\documentclass{aa}

\usepackage{graphicx}
\usepackage{amssymb}
\usepackage{natbib}
\usepackage{epsf}
\usepackage{txfonts}

\newcommand{\lam}{$\lambda$}

\newcommand{\ecss} {erg~cm$^{-2}$~s$^{-1}$~sr$^{-1}$} 
\newcommand{\ecsa} {erg cm$^{-2}$ s$^{-1}$~sr$^{-1}$~\AA$^{-1}$} 
\newcommand{\kms} {km\ s$^{-1}$} 
\newcommand{\deltE}{\Delta\kern-1ptE}

\newcommand{\hinode}{\emph{Hinode}}

\newcommand{\as}{$^{\prime\prime}$}
\newcommand{\ep}{EIS\_PREP}
\newcommand{\diag}{DIAG\_40X180\_S0\_30S}

\title{High-precision density measurements in the solar corona: I.~Analysis methods and results for Fe XII and Fe XIII}

\author{P.R.\ Young\inst{1,2,3}\thanks{Present address: Space Science Division,
            Naval Research Laboratory, Washington, DC 20375, U.S.A.}
\and
T.\ Watanabe\inst{4}
\and
H.\ Hara\inst{4}
\and
J.T.\ Mariska\inst{3}
}

\institute{Rutherford Appleton Laboratory, Chilton, Didcot,
  Oxfordshire, OX11 0QX, U.K.
\and
George Mason University, 4400 University Drive, Fairfax, VA 22030, U.S.A.
\and
Space Science Division, Naval Research Laboratory,
  Washington, DC 20375, U.S.A.
\and
National Astronomical Observatory of Japan, National
  Institutes of Natural Sciences, Mitaka, Tokyo, 181-8588, Japan} 

\date{Received / Accepted}

\abstract
{}
{The  EUV Imaging Spectrometer (EIS) instrument on
  board the \emph{Hinode} satellite has access to some of the
  best coronal density diagnostics, and  the high
  sensitivity of the instrument now allows electron number density,
  $N_{\rm e}$,  measurements to an
  unprecedented precision of up to $\pm 5$~\%\ in active regions.
This
  paper gives a thorough overview of data analysis issues for the best
  diagnostics of \ion{Fe}{xii} and \ion{Fe}{xiii} and assesses the
  accuracy of the measurements.}
{Two density diagnostics each from \ion{Fe}{xii}
  (\lam186.88/\lam195.12 and \lam196.64/\lam195.12) and \ion{Fe}{xiii}
  (\lam196.54/\lam202.04 and \lam203.82/\lam202.04)
  are analysed in two active region datasets from 2007 May 3 and 6 that yield densities in
  the range $8.5\le \log\,(N_{\rm e}/{\rm cm}^{-3})\le 11.0$. The
  densities are derived using v5.2 of the CHIANTI atomic
  database. Blending, line fitting, and instrumental issues are
  discussed, and line fit parameters presented.}
{The \ion{Fe}{xii} and \ion{Fe}{xiii} diagnostics show broadly the
  same trend in density across the active region, consistent with
  their similar temperatures of formation.
However, the high precision of the EIS
  measurements demonstrates significant discrepancies of up to 0.5~dex
in derived $\log\,N_{\rm e}$ values, with \ion{Fe}{xii} always giving
higher densities than \ion{Fe}{xiii}. The discrepancies may partly be
due to real physical differences between the emitting regions of the
two plasmas, but the dominant factor lies in the atomic models of the
two ions.  Two specific problems are
identified for \ion{Fe}{xii} \lam196.64 and \ion{Fe}{xiii} \lam203.82:
the former is found to be underestimated in strength by the CHIANTI
atomic model, while the high-density limit of the
\lam203.82/\lam202.04 ratio appears to be inaccurate in the CHIANTI
atomic model. The small grating tilt of the EIS instrument is found to be
very significant when deriving densities from emission lines separated
by more than a few angstroms. Revised wavelengths of $196.518\pm 0.003$~\AA\ and
$196.647\pm 0.003$~\AA\ are suggested for the \ion{Fe}{xiii} \lam196.54 and
\ion{Fe}{xii} \lam196.64 lines, respectively.}
{}

\begin{document}

\titlerunning{High precision density measurements in the solar
  corona I}
\authorrunning{Young et al.}

\maketitle

\section{Introduction}

Density is a fundamental plasma parameter that is  accessible to astrophysical
spectroscopists through the \emph{density diagnostic} -- a ratio of two
emission lines that is 
sensitive to the electron density of the emitting plasma.
High accuracy density measurements are of direct importance for
constraining energy and pressure
balance within coronal structures, and also of indirect importance to
the measure of the `filling factor' of the plasma, i.e., the fraction
(by volume) of the observed plasma structure that is actually emitting
radiation. The latter is of significance when the spatial resolution
of the telescope is not sufficient to resolve the physical structures
of interest, as is often the case even for our nearest star.
The power of emission line density diagnostics lies in the fact that,
when the plasma is optically thin, the derived density depends purely
on the atomic data for the emitting ion and is  free of
any physical assumptions about the volume, abundances or temperature
structure within the plasma.

Density diagnostics for use in both analyses of stellar coronae and
nebulae have been described and exploited for many years. For the
solar corona, reviews of early work are provided by \citet{dere81} and
\citet{mariska92}. Steady improvements in the accuracy of atomic data
calculations and access to high resolution, calibrated ultraviolet
spectra from 
space-borne instruments have led to a number of detailed studies of
density diagnostics from both individual ions and sequences of ions,
e.g., \citet{brickhouse95}, \citet{laming97} and \citet{young98}.

A particular advance has been the routine access to most of the
wavelength range 150--1600~\AA\ by the CDS and SUMER instruments on board
the \emph{SOHO} spacecraft for over a 12 year period. A wide range of density
diagnostics from many different ions 
has been exploited \citep[e.g.,][]{doschek97,laming97,young99,delzanna03},
however the density measurements from
these two instruments are generally accurate only to the 30--40~\%\ level at
best and often it is necessary to bin multiple spatial pixels or
exposures to
obtain a useful signal. This is because the most commonly-used
diagnostics generally involve only weak-to-medium strength emission
lines and/or the diagnostics themselves do not have high sensitivity
to density.

The best density diagnostics at typical coronal temperatures
(1--3~million~K) belong to the set of iron ions \ion{Fe}{ix--xv}. They
arise because of the complex atomic structures of these ions which
lead to a number of metastable levels that are the source of the
density sensitivity of the emission line ratios \citep[e.g.,
Sect. 6.3.1 of][]{dere81}. The density diagnostics of these iron ions
are mostly found below 300~\AA, a wavelength region not well covered
by the \emph{SOHO} spectrometers.

The EUV Imaging Spectromer (EIS) on board the \hinode\
satellite -- launched during 2006 September -- observes two wavelength 
ranges below 300~\AA\, and, for the first time, scientists have routine
access to high spectral and spatial resolution data in this region.
A key factor is the high
sensitivity of the instrument which is enabled by a simple optical
design that incorporates multilayer coatings \citep{culhane07}. The
sensitivity is particularly high in the range 185--205~\AA\ where are
found excellent density diagnostics from \ion{Fe}{xii} and \ion{Fe}{xiii}.
As
demonstrated in the following sections, diagnostics from these ions
yield density measurements with an unprecedented 
precision of up to 5~\%\ in \emph{individual} spatial pixels at
exposure times of 30~s or less -- a vast improvement over earlier
instruments. 

The diagnostic potential of \ion{Fe}{xii} and \ion{Fe}{xiii} has been
recognised for many years. The first sophisticated atomic calculation
for \ion{Fe}{xii} was performed by \citet{flower77} and a number of UV
density diagnostics have been studied
\citep[e.g.,][]{feldman83,tayal91}, but for many years 
there were large discrepancies between predictions from atomic data
models and measured 
line intensities which were summarised in \citet{binello01}. Finally
the large $R$-matrix calculation performed by \citet{storey05}
resolved many of the problems \citep{delzanna05}. The three emission
lines used in the present work (\lam186.88, \lam195.12 and \lam196.64)
were shown by \citet{delzanna05} to give good
agreement with other line 
ratios from \ion{Fe}{xii}, but each line is affected by line
blending. This will be discussed in detail during the course of the
present work.

For \ion{Fe}{xiii}, \citet{flower74}
calculated atomic data and produced line ratio calculations that
yielded reasonably consistent density measurements from a quiet Sun
spectrum. Detailed comparisons of the more recent atomic data models of
\ion{Fe}{xiii} in the CHIANTI atomic database \citep{dere97} with
spectra from the SERTS rocket flight experiments \citep{neupert92} have
been performed by \citet{young98} and \citet{landi02}. In particular,
the latter work found that the \lam196.54/\lam202.04 and
\lam203.82/\lam202.04 ratios  used in the present work give
density estimates in good agreement with each other and also with
other ions formed at similar temperatures.

Early density results from EIS obtained using the \ion{Fe}{xii} and
\ion{Fe}{xiii} diagnostics have been presented in
\citet{brooks07}, \citet{watanabe07}, \citet{doschek07a},
\citet{doschek07b} and \citet{dere07}. In particular we highlight the
density maps shown in \citet{doschek07b} that demonstrate the high
quality of the EIS density measurements, and the density comparisons shown
in \citet{watanabe07} that show excellent agreement
between four different iron ion density diagnostics.

In this paper density measurements from two diagnostics each of the
\ion{Fe}{xii} and \ion{Fe}{xiii}  ions are presented from two active
region data sets. A careful discussion of line fitting and blending
issues is given, and an estimate of the accuracy 
of the density measurements is made. A following paper will use the
derived densities to estimate column depths and filling factors for
the two ions.

\begin{table*}[h]
\caption{Transition information and CHIANTI wavelengths for the
  density diagnostics}
\begin{center}
\begin{tabular}{llll}
\noalign{\hrule}
\noalign{\smallskip}
\noalign{\hrule}
\noalign{\smallskip}
Ion & Wavelength/\AA & Transition & EA$^a$/cm$^{-2}$ \\
\noalign{\hrule}
\noalign{\smallskip}
\ion{Fe}{xii}
     &186.854 &$3s^23p^3$ $^2D_{3/2}$ -- $3s^23p^2(^3P)3d$ $^2F_{5/2}$
     &0.114 \\
     &186.887 &$3s^23p^3$ $^2D_{5/2}$ -- $3s^23p^2(^3P)3d$ $^2F_{7/2}$
     &0.115 \\
     &195.119 &$3s^23p^3$ $^4S_{3/2}$ -- $3s^23p^2(^3P)3d$ $^4P_{5/2}$ 
     &0.303 \\
     &196.640 &$3s^23p^3$ $^2D_{5/2}$ -- $3s3p^3(^1D)3d$ $^2D_{5/2}$ 
     &0.301 \\
     
\\
\ion{Fe}{xiii} 
     &196.540 &$3s^23p^2$ $^1D_2$ -- $3s^23p3d$ $^1F_3$ & 0.302 \\
     &202.044 &$3s^23p^2$ $^3P_0$ -- $3s^23p3d$ $^3P_1$ & 0.084 \\
     &203.797 &$3s^23p^2$ $^3P_2$ -- $3s^23p3d$ $^3D_2$ & 0.044 \\
     &203.828 &$3s^23p^2$ $^3P_2$ -- $3s^23p3d$ $^3D_3$ & 0.044 \\
\noalign{\hrule}
\noalign{\smallskip}
\multicolumn{4}{p{6cm}}{$^a$ Effective area values.}\\
\end{tabular}
\end{center}
\label{tbl.lines}
\end{table*}

\section{Instrument details}

The EUV Imaging Spectrometer (EIS) on \emph{Hinode} observes two
wavelength ranges: 170--212~\AA\ and 246--292~\AA. The instrument is
described in detail by \citet{korendyke06} and \citet{culhane07}. The key
optical elements are: a parabolic primary mirror for focussing incoming
radiation; a uniform line-spaced grating for dispersing the radiation
and two back-thinned, EUV sensitive CCDs for detecting the
radiation. Two aluminium filters lie on the optical path for blocking
visible radiation, and a slit-slot assembly is situated between the
mirror and grating containing four different-sized slits for selecting
the spatial region to be sent to the spectromer. The optical surfaces
of both the mirror and grating are divided in two and coated with
multilayer coatings optimised for the two wavelength bands.  Of the
four entrance slits, two are available for detailed
spectroscopy work with projected widths at the Sun of
1\as\ and 2\as. Pixel sizes on the detector are 1\as\ (spatial
dimension) and 0.0223~\AA\ (spectral dimension). For single exposures
the maximum field-of-view in Solar-Y is 512\as.

For density measurements, the absolute radiometric calibration of the
EIS instrument is not important, only the relative calibration between
the wavelengths of the lines. The key uncertainty is whether the
relative reflectivities from the mirror and grating between two
wavelengths deviate from the values contained in the calibration
pipeline. There are two components: the uncertainty due to variations in
reflectivity across the optical surfaces; and a systematic error in
the reflectivity curves due to some change in the instrument between
the laboratory calibration measurements and launch.
For the former, measurements performed on flight-quality mirrors and gratings by
\citet{seely04} revealed that, when a beamline is scanned
across the optical surfaces, the reflectivity as a function of wavelength is
found to vary. \citet{seely04} averaged the resulting reflectivity
curves and the standard deviations of the data points for the
wavelengths considered here range from 3 to 8~\%. In orbit the optical
surfaces are fully illuminated which will serve to average out the
variations and so these standard deviation values will be over-estimates of
the actual uncertainty in the relative calibration. In addition, these
percentage errors are for the absolute reflectivity -- the standard
deviation for the relative reflectivities between two lines close in
wavelength (e.g., the \ion{Fe}{xiii} \lam202.04 and \lam203.82) will
likely be significantly less. For this reason we
do not include any estimate of the relative calibration uncertainty in
the error budget for derived densities. The error bars for
the density measurements presented here are thus underestimates,
but we believe the uncertainties in the relative calibration are
comparable to or smaller than these errors and would not significantly
impact the conclusions of this work. Of greater concern for the
present work is whether there is a systematic error in the shape of
the multilayer reflectivity curves used in the EIS calibration. This
can be investigated through the study of emission line ratios that are
insensitive to the plasma conditions -- an example being the
\ion{Fe}{xii} ratios considered in Appendix~\ref{app.fe12}. The
question of systematic errors is considered in the later sections when
the density results are presented.

The basic unit of observation for EIS is a \emph{raster} which is a
consecutive set of exposures that either scan a spatial region (from
solar-west to solar-east) on the
Sun, or are fixed relative to the Sun's surface (``sit-and-stare''
observations). A \emph{study} is defined as a collection of one or
more rasters
that is designed to perform a particular science task. Each study and
raster has a unique 20 character acronym that is used to identify it
for science planning purposes.

On account of data download volume restrictions, complete CCD images
from each EIS exposure are not routinely transmitted to Earth.
A key part of the
raster definition process is thus to define `wavelength windows' on the
detector for selecting particular emission lines and/or a reduced
spatial coverage in the Solar-Y direction. The widths of the windows
need to be chosen carefully by the scientist so that accurate fitting
of the emission line profiles will be possible. Typically window
widths are chosen to be 24--32 pixels (0.54--0.71~\AA) but for some of
the emission lines considered here wider windows are necessary, and
this will be discussed in the following sections.

\section{Atomic data}

To convert a measured line ratio value to a density value it is
necessary to have a model for the atomic processes within the ion that
predicts the relative strengths of the emission lines as a function of
density. Here we use version 5.2 of the CHIANTI atomic
database \citep{landi06,dere97} which contains atomic data and
analysis software for calculating line ratios as a function of density.
For \ion{Fe}{xii} the data consist of
energy levels and radiative decay rates from \citet{delzanna05},
electron collision strengths 
from \citet{storey05}, and proton collision rate coefficients from
\citet{landman78}. For \ion{Fe}{xiii} energy levels are from
\citet{penn94}, \citet{jupen93} and version 1.0 of the NIST database;
radiative decay rates are from \citet{young04}, electron collision
strengths from \citet{gupta98}, and proton collision rate coefficients
from \citet{landman75}. During the course of the present work, the
CHIANTI team found an
error in the CHIANTI electron collision strength file for
\ion{Fe}{xiii}, which actually contained older data from
\citet{fawcett89} rather than the data of \citet{gupta98} that were supposed to
have been added in version~3 of the database \citep{dere01}. With the
\citet{gupta98} 
dataset \ion{Fe}{xiii} yields significantly different densities
\citep{landi02}. 

The atomic transitions of the \ion{Fe}{xii} and \ion{Fe}{xiii} density
diagnostic lines considered here are given in
Table~\ref{tbl.lines}, and plots showing how the ratios are predicted
to vary with density are shown in Fig.~3 of \citet{young07b}. The
diagnostics themselves are the three 
recommended by \citet{young07b} -- \ion{Fe}{xii}
\lam196.64/\lam195.12, and \ion{Fe}{xiii} \lam196.54/\lam202.04 and
\lam203.82/\lam202.04 -- together with \ion{Fe}{xii}
\lam186.88/\lam195.12 which is comparable to \lam196.64/\lam195.12 but
concern was expressed by \citet{young07b} over the blending line
\ion{S}{xi} \lam186.84. This issue will be discussed further in
Sect.~\ref{sect.186}. Between them, these four density diagnostics are
the best available in the EIS wavelength bands. The \ion{Fe}{xii}
diagnostics benefit from the high instrument sensitivity at
195.12~\AA\ which leads to accurate measurements of the line
ratios. The \ion{Fe}{xii} diagnostics also benefit from a wide sensitivity to
density from $10^8$ to 10$^{12}$~cm$^{-3}$ that arises through the shift
of population from the ground $^4S$ term to first the $^2D$
term and then the $^2P$ term as density increases over this range. The
EIS sensitivity for \ion{Fe}{xiii} \lam202.04 and \lam203.82 is lower
than for the \ion{Fe}{xii} lines (Table~\ref{tbl.lines}), but the
value of the diagnostics 
lies in the strong sensitivity of the lines to density. E.g., the
\lam203.82/\lam202.04 ratio varies by a factor 43 from $10^8$ to
10$^{10}$~cm$^{-3}$, whereas \ion{Fe}{xii} \lam196.64/\lam195.12
varies by a factor 8. This high sensitivity means that the error bars
on a measured ratio translate to relatively small errors on the derived
density. 

\begin{figure}[h]
\centerline{\epsfxsize=9cm\epsfbox{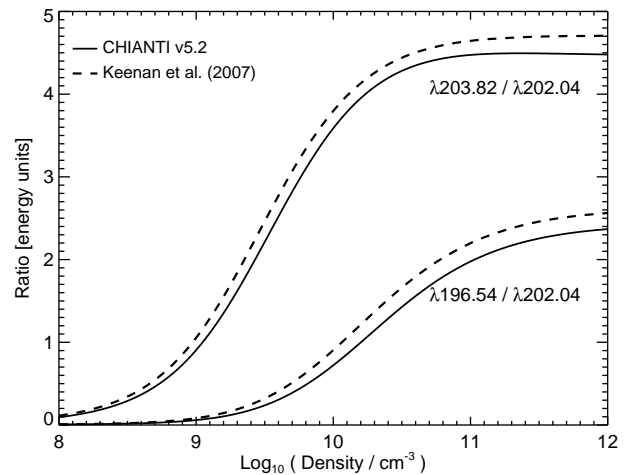}}
\caption{Theoretical variation of emission line ratios versus electron
  density for \ion{Fe}{xiii} \lam203.82/\lam202.04 and
  \lam196.54/\lam202.04. The solid lines show the results from the
  modified CHIANTI v5.2 model while the dashed lines show results from
  the \ion{Fe}{xiii} model of \citet{keenan07}.}
\label{fig.aggarwal}
\end{figure}

A more recent set of collision  data for \ion{Fe}{xiii} than that in
CHIANTI was published by \citet{aggarwal05}, and these have been
compared with the v5.2 CHIANTI \ion{Fe}{xiii} model and SERTS rocket
flight observations in \citet{keenan07}. The \ion{Fe}{xiii} model used
by \citet{keenan07} consisted of collision  data from \citet{aggarwal05} and
radiative data from \citet{aggarwal04}. The agreement with the
CHIANTI model was found to be very good (although this was the CHIANTI
model containing the Fawcett \& Mason, 1989, data rather than the
model used here). Fig.~\ref{fig.aggarwal} compares the theoretical
ratios from our revised CHIANTI model and \citet{keenan07} for the two \ion{Fe}{xiii}
ratios. In both cases, for a given ratio value, the \citet{aggarwal05}
data yield lower densities by around 0.1--0.2~dex. The high-density
limits from \citet{keenan07} are higher than those from CHIANTI
which has a significant impact on derived results from high density
plasmas. These issues will be discussed further in
Sects.~\ref{sect.may6} and \ref{sect.may3}.   \citet{keenan07}
presented high resolution spectra from the SERTS-95 rocket flight
\citep{brosius98} where they were able to resolve the two components
of the \ion{Fe}{xiii} \lam203.82 blend, i.e., the lines at 203.79 and
203.83~\AA. This is not possible with the lower resolution EIS
spectra. \citet{keenan07} found that the \lam203.79/\lam203.83 ratio
was significantly different from theoretical predictions, with an
observed value of 0.20 and a theoretical value of 0.32. This raises
concerns about the suitability of using the blended \lam203.82 line as
part of a density diagnostic. We note, however, that summing the
\lam203.79 and \lam203.83 intensities from the SERTS-95 spectrum and
forming a density diagnostic with \lam202.04, the derived density
\citep[using the][model]{keenan07} is $\log\,N_{\rm e}=9.1$ which is
consistent with the other \ion{Fe}{xiii} densities presented in  Table~9 of
\citet{keenan07}. We note the line widths measured by
\citet{keenan07} for \lam203.79 and \lam203.83 are significantly
different which is not expected from two lines of the same ion,
and this may partly be responsible for the discrepancy between theory
and observation for their intensity ratio.  For the remainder of this
work we will concentrate on the density results from the modified
\ion{Fe}{xiii} CHIANTI model as CHIANTI is widely used amongst the
solar physics community.

The standard CHIANTI models for \ion{Fe}{xii} and \ion{Fe}{xiii}
include electron excitation/de-excitation, spontaneous radiative decay
and proton excitation/de-excitation. Further processes that can be
included are photo-excitation and stimulated emission which
can be significant in low density conditions when electron collisions
are less frequent. The method for including these latter processes in
CHIANTI is described in \citet{young03}, and examples of the effects
of the processes on density diagnostics are provided in
\citet{flower73} and \citet{young99a,young99b}. For the solar case,
the radiation field to be
considered is the photospheric continuum which is modelled in CHIANTI
as a black-body. For \ion{Fe}{xii}, such a radiation field has no
effect on the density sensitive ratios considered in this work. For
\ion{Fe}{xiii} there is no significant effect on
\lam196.64/\lam202.04, but there is an effect for
\lam203.82/\lam202.04.

\begin{figure}[h]
\centerline{\epsfxsize=9cm\epsfbox{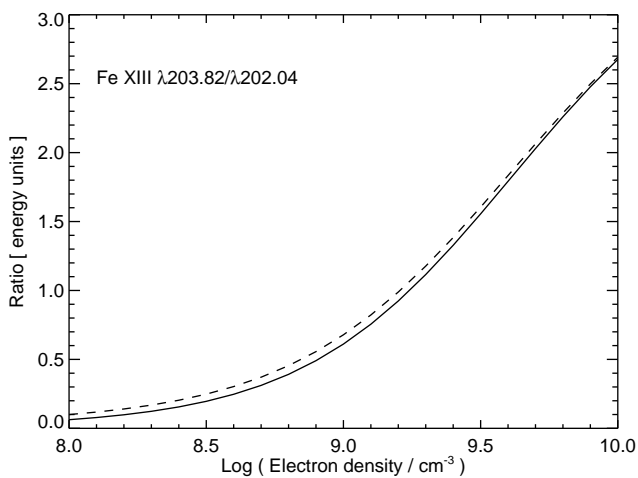}}
\caption{The theoretical variation of the \ion{Fe}{xiii}
  \lam203.82/\lam202.04 ratio as a function of density, as derived
  from CHIANTI. Two cases are shown: the curve calculated assuming no
  incident radiation field (solid line), and the curve calculated
  assuming the \ion{Fe}{xiii} emitting plasma is located 10\as\ above
  a photosphere of temperature 6000~K.}
\label{fig.fe13-photoexc}
\end{figure}

Fig.~\ref{fig.fe13-photoexc} compares the predicted variation of
\ion{Fe}{xiii} \lam203.82/\lam202.04 with density for two cases: when
there is no incident radiation field (solid line) and when the
\ion{Fe}{xiii} emitting plasma is located 10\as\ above the solar
surface (dashed line). The radiation field serves to push the curve
towards lower densities. A measured intensity ratio will thus yield a
lower density when the radiation field is taken into account: for
$\log\,N_{\rm e} \approx 9$ the reduction is about 0.05~dex. The change
becomes less for higher densities, being negligible above
$\log\,N_{\rm e}=10$. The location of the \ion{Fe}{xiii} emitting
plasma above the photosphere is also a factor, with greater distances
leading to a smaller effect. For the present work the effects of an
incident radiation field will not be considered any further since the
magnitude of the effect is small compared to some of the discrepancies
discussed later, and also it is not possible to make a general
statement about the location of the \ion{Fe}{xiii} emitting
plasma in the data sets considered.

The present work is focussed on density diagnostics and so wavelengths
and velocity measurements are not important. For all reference
wavelengths quoted in the text, we use the the values from v.5.2 of
the CHIANTI database \citep{landi06}. In particular we note the
revised \ion{Fe}{xii} wavelengths of \citet{delzanna05} which are
included in CHIANTI and are valuable for interpreting the
\ion{Fe}{xii} \lam195.12 line (Sect.~\ref{sect.195}). A by-product of
the Gaussian fits to the emission line profiles is line centroid
measurements for each line, and these can be used to assess the
accuracy of the CHIANTI wavelengths (Sect.~\ref{sect.integrity}).

\section{Data calibration and error bars}

The EIS data presented here were all calibrated from the level-0 FITS
files using the EIS\_PREP routine available in the \emph{Solarsoft}\footnote{\emph{Solarsoft} is
  a set of  integrated software libraries, databases, and system
  utilities that provide a common programming and data analysis
  environment for Solar Physics. It is available at
  http://www.lmsal.com/solarsoft.}
IDL distribution. In addition to converting the measured CCD signal
into calibrated intensity units, a key part of \ep\ is to flag bad data
points. These can arise through pixel saturation, cosmic ray hits, or
simply defective pixels on the CCD. In this work all such pixels were
marked as `missing' data and not included in the data analysis.
The
central outputs of EIS\_PREP are two level-1 FITS files, one
containing calibrated 
intensities at each pixel, and the other containing error bars on
these intensities. The steps performed by the current version of \ep\
at the time of this analysis are described below in detail.

The first step of  EIS\_PREP is to flag any saturated data. The EIS
CCDs have a 14 bit dynamic range and so saturation occurs at
16,383~data numbers (DN). All such pixels are flagged as missing as described above.

In the raw data, the spectra are found to sit on a background of
around 500~DN that arises principally from the CCD bias, and secondly
from the CCD dark current. It is not possible to estimate the CCD bias
level directly for EIS data, so the bias and dark current levels are
estimated directly from the science data as follows. For each 3D data window 
2~\%\ of the detector pixels are isolated that have the lowest DN
values. The median DN value of these 2~\%\ pixels is then set to be
the background level and it is subtracted from the DN values of each
pixel. 

Anomalously bright pixels are found on the EIS CCD images that arise
from `hot pixels', `warm pixels' and cosmic rays. 
Both hot pixels and warm pixels are single pixels that have
anomalously high DN values. A hot pixel is defined to be one that yields
25,000~electrons~pixel$^{-1}$~s$^{-1}$ at room temperature (a
specification from the CCD manufacturer). Pixels
that fall below this threshold but are still clearly identified as
being anomalous when inspecting the data are referred to as `warm'
pixels. Maps of the locations of hot pixels are generated by the EIS
team every 2--4 
weeks following 
inspection of 100~s dark exposures and they are stored in
\emph{Solarsoft}. The hot pixel map that is closest in time to the
science observation is used by EIS\_PREP to mark the hot pixels as
missing data.

Before removing warm pixels and cosmic rays, the next step for EIS\_PREP is to flag
the pixels affected by dust on the CCD. Several small pieces of dust
accumulated on the CCD before launch and are found to completely block
the solar signal on the CCD at their locations. They are fixed in
position and cover less 
than 0.1~\%\ of the CCD, however two of the pieces do affect the
strong lines \ion{Fe}{xi} \lam\lam188.23, 188.30 and \ion{Fe}{xii}
\lam193.51 such that the lines can not be used over 15--30\as\
spatial ranges in solar-Y.

At the time of
performing the present analysis, EIS\_PREP did not specifically remove
warm pixels, but many of these were removed naturally by the cosmic
ray removal routine, EIS\_DESPIKE.
This latter is a wrapper routine that calls NEW\_SPIKE, a routine
developed  for removing cosmic rays from \emph{SOHO}/Coronal
Diagnostic Spectrometer (CDS) data sets \citep{thompson98,
pike00}. 
For CDS data processing
it was typical for not only the identified CCD pixels to be flagged,
but also the nearest-neighbour pixels on the CCD. This is because there is
often residual signal from the cosmic ray next to the brightest
pixels. EIS sees significantly less cosmic rays than CDS apart from
during the $\approx$5 minute passes through the South Atlantic
Anomaly, and the most useful function of EIS\_DESPIKE is actually to flag
warm pixels. Since warm pixels are only single pixel events, then the
nearest-neighbour option is switched off for EIS. It is to be noted
that the NEW\_SPIKE routine was designed to be cautious when removing
cosmic rays from line profiles thus many weak warm pixels found within
spectral lines are not removed, artificially enhancing the
emission line intensities at these locations. For the 2007 May
data sets analysed in the present work only around 2~\%\ of the CCD
pixels are warm pixels and so this is not a significant problem.

The
final step of \ep\ is to convert DN values into intensities  in units
\ecsa. The errors on the intensities are computed assuming photon
statistics together with an error estimate of the dark current of
2.5~DN. 

\subsection{CCD spatial offsets}

A further instrumental effect accounted for in the present work is the
spatial offsets between different wavelengths. Spatial offsets between
images formed from lines on the two different EIS CCDs have been
discussed by \citet{young07a}. The emission lines discussed here are
exclusively from the short wavelength band and so these latter spatial
offsets do not apply. However, the spectra on the short wavelength (SW) CCD are slightly
tilted relative to the CCD's axes such that images from short
wavelength lines are 
marginally higher on the CCD than long wavelength
lines. The effect is due to a slight misalignment of the EIS grating
relative to the CCD and which we refer to here as the ``grating
tilt''. Note that this is not to be confused with the ``slit tilt''
\citep[e.g.,][]{mariska07} which is a misalignment of the EIS slits
relative to the CCD. This latter effect results in slit images being
slighly tilted on the detector, but does not result in image
misalignment between lines at different wavelengths.

The grating tilt is discussed in more detail in
Appendix~\ref{sect.offsets} where the gradient of the dispersion axis
on the CCD is found to be $-0.0792$~pixels/\AA\ for the EIS SW band.
Despite this small
value, the effect on line ratios for lines separated by several
angstroms can be significant as demonstrated in
Appendix~\ref{sect.offsets}.
For all of the density diagnostic lines considered here
except \ion{Fe}{xii} \lam195.12 the 2D intensity arrays that arise
after Gaussian fitting has been performed have been adjusted to
account for the spatial offsets. The intensities have been modified
according to Eq.~\ref{eq.offset}, and the intensity errors have also
been modified accordingly.

\section{Line fitting -- method and blending issues}\label{sect.lines}

In the following sections we describe the methods used to
automatically fit the density diagnostic emission lines with
Gaussians. Firstly, though, we must ask two questions: is it necessary
to fit Gaussians to 
the data? and is the Gaussian an appropriate function to use?  For
density diagnostic work only the line intensity is needed, so
it is simply necessary to sum up the intensity from the wavelength
bins containing the emission lines without the need to resort to a
model for the line profile. However, as will be seen, some of the
lines are blended with other lines and so it is essential to have a
model for the line profile in order to resolve the blending
components. Fitting the lines with Gaussians yields the line centroid
and line width in addition to the intensity, which thus allow the
density to be related to velocity and non-thermal broadening
parameters of the solar plasma, providing a much more powerful
data set for scientific analysis. The widths and velocities are also
valuable for assessing the accuracy of the line fitting as will be
demonstrated in Sect.~\ref{sect.integrity}. 

As to the appropriateness of Gaussian fits, in normal conditions the
EIS line profiles are found to be very well represented by a Gaussian
function. In particular, no evidence is found for asymmetries  or
enhanced wings in 
the line spread function. However, non-Gaussian profiles due to real
solar events are seen
often with EIS 
and examples are given in \citet{imada07} and
\citet{chifor08}. Typically such events are isolated and can be readily
identified following an automatic Gaussian fit to the data, as they
will have anomalous line widths and/or velocities. For the data
analysed in the present work there is no evidence for any significant
deviations from Gaussian profiles.

The high quality of the EIS data mean that the lines considered here
can be automatically fit with Gaussians at virtually every spatial
pixel in the EIS images. However, care has to be taken in applying
automatic fitting routines due to the influence of both nearby and
blending emission lines. In this section we go through each line and
highlight blending and fitting issues. A key point is that the fits
must be accurate and robust for the wide range of conditions
encountered in the solar atmosphere. The discussion in this section
builds on the 
initial overview of the lines presented in \citet{young07b}.

To fit Gaussians to the EIS lines 
the MPFIT procedures of
C.~Markwardt\footnote{http://astrog.physics.wisc.edu/$\sim$craigm/idl/idl.html.}
have been implemented to automatically fit
the calibrated line intensities from the spectral line windows output
by the IDL routine EIS\_GETWINDATA. The spectrum from each image pixel
is fitted, with data points weighted with the 1$\sigma$ error bars
calculated by EIS\_PREP. The routine developed by P.R.~Young for
fitting single Gaussians to the 
spectral lines is called EIS\_AUTO\_FIT and is available in the
\emph{Solarsoft} IDL distribution. For some of the spectral features
discussed below it was necessary to perform multiple Gaussian fits
and custom versions of EIS\_AUTO\_FIT were created for these.

In the sections that follow the reference wavelengths of emission
lines (denoted by, e.g., \lam195.12) are taken from v5.2 of the
CHIANTI atomic database. Wavelengths are given to two significant
figures, except where greater accuracy is required. Two of the density
diagnostic lines are actually self-blends of two transitions from the
ions and we use \lam186.88 to denote the blend of \ion{Fe}{xii}
\lam186.854 and \lam186.887, and \lam203.82 to denote the blend of
\ion{Fe}{xiii} \lam203.797 and \lam203.828.

\subsection{Fe\,XII 186.88}\label{sect.186}

\begin{figure}[h]
\centerline{\epsfxsize=9cm\epsfbox{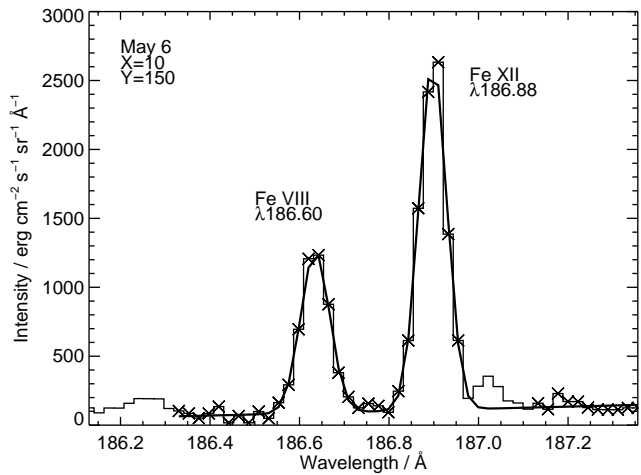}}
\caption{An example double-Gaussian fit to the \ion{Fe}{viii} \lam186.60 and
  \ion{Fe}{xii} \lam186.88 lines from the May 6 data set. The chosen
  spatial pixel is indicated. The data are
  plotted as a thin line and 
  the fit with a thick line. The wavelength pixels used in the fitting are
  indicated by crosses. Note that the weak line on the long wavelength
  side of \ion{Fe}{xii} \lam186.88 has not been included in the fit.}
\label{fig.186-example}
\end{figure}

\begin{figure}[h]
\centerline{\epsfxsize=9cm\epsfbox{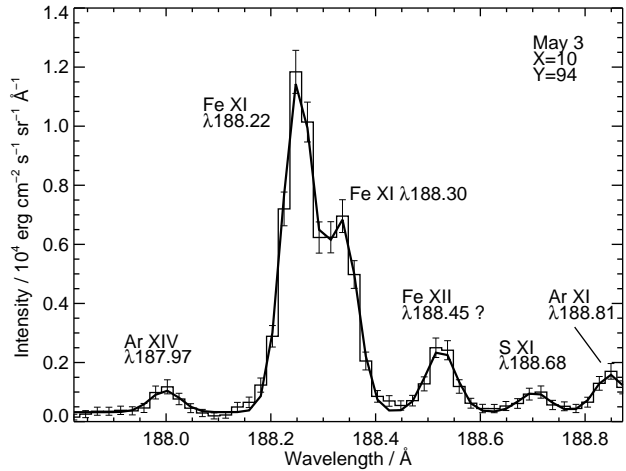}}
\caption{Spectrum from a single spatial pixel in the May 3 dataset
  showing the wavelength window used to obtain the \ion{Fe}{xi}
  \lam\lam188.22, 188.30 feature. A six Gaussian fit to all of the lines in
  the window is shown. The weak \ion{S}{xi} \lam188.68 line used to
  estimate the contribution of \ion{S}{xi} to \ion{Fe}{xii} \lam186.88
  is indicated.}
\label{fig.fe11-s11}
\end{figure}

The emission line at 186.88~\AA\ is generally strong and well-resolved
in active regions (Fig.~\ref{fig.186-example}) and it largely
comprises of  two \ion{Fe}{xii} transitions
with CHIANTI wavelengths of 186.854 and 186.887~\AA. \ion{S}{xi}
\lam186.839  is a known blend and its contribution can be assessed
from the \lam188.617 or \lam191.266 transitions which are related
by branching ratios as they arise from the same upper level ($2s2p^3$
$^3S_1$). From CHIANTI the \lam186.84, \lam188.68 and \lam191.27 intensities
lie in the ratios 0.195:0.580:1.0. In the data sets discussed in this
work 
\lam191.27 is not observed, but \lam188.68 lies in the window
containing the strong \ion{Fe}{xi} lines at 188.22 and 188.30~\AA\ (Fig.~\ref{fig.fe11-s11}).

\begin{figure*}[h]
\centerline{\epsfxsize=14cm\epsfbox{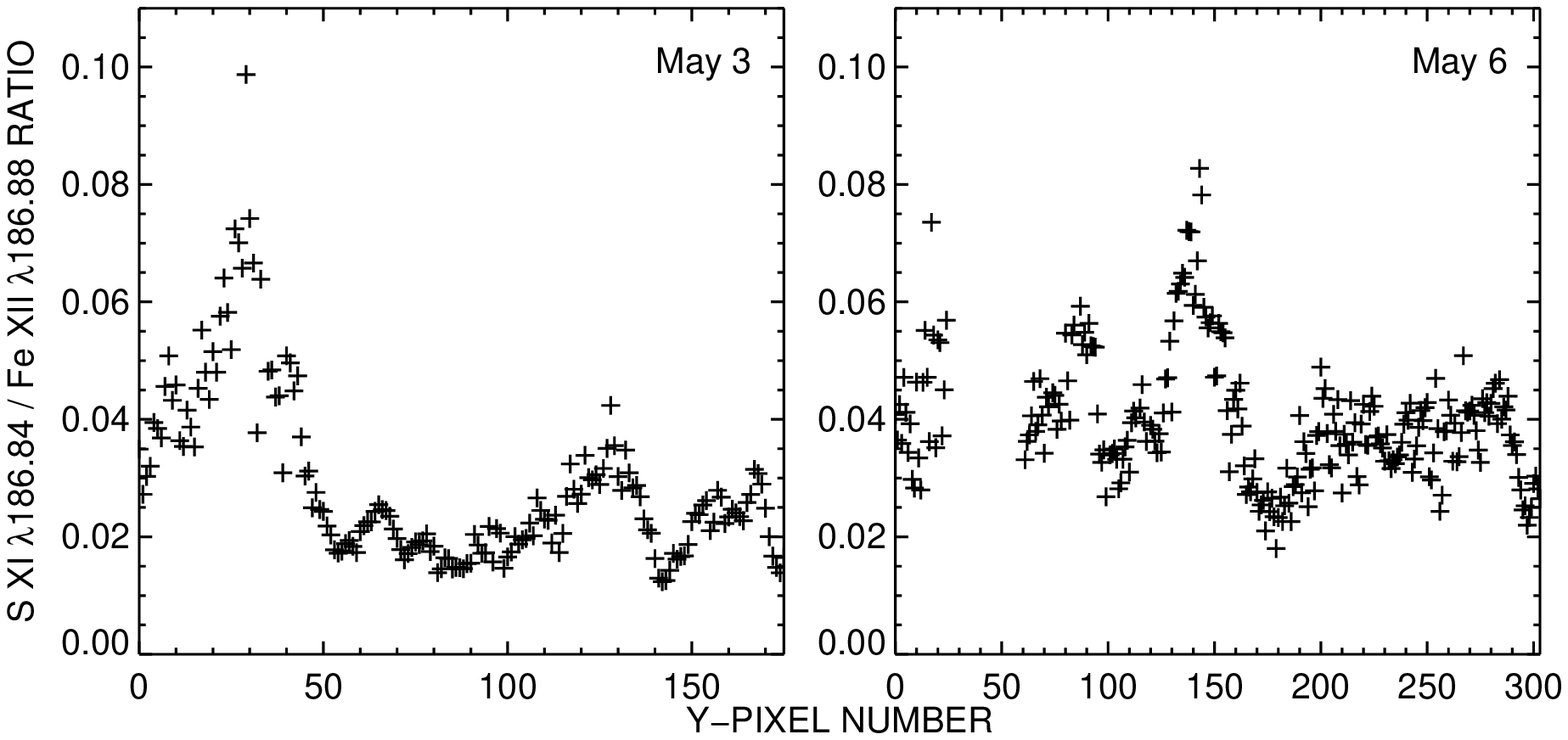}}
\caption{The intensity ratio of \ion{S}{xi} \lam188.64 to the blended
  \ion{Fe}{xi} \lam186.88 feature for two columns of spatial pixels in 
  the May 3 (left panel) and May 6 (right panel) data sets. The
  \lam188.64 has been estimated from fits to \ion{S}{xi}
  \lam188.68. For the May~6 data set, accurate fits to \lam188.68 were not
  possible due to a dust particle on the CCD at this position.}
\label{fig.s11-fe11}
\end{figure*}

To demonstrate the contribution of \ion{S}{xi} to the \ion{Fe}{xii}
\lam186.88 line the intensity of \ion{S}{xi} \lam188.68 in the May 3
and May 6 data sets considered in
Sects.~\ref{sect.may6} 
and \ref{sect.may3} has been measured in the \ion{Fe}{xi} \lam188.22
window. Fig.~\ref{fig.fe11-s11} shows the \lam188.22 wavelength
window, where a total of six emission lines can be seen. A six
Gaussian fit was applied and, to ensure robust fits, each line was
assumed to have the same width. Otherwise amplitudes and centroids
were free to vary. In both the May 3 and May 6 active region datasets,
the six Gaussian fits were generally accurate. A 35 pixel high
region on the detector  is
affected by the presence of a dust particle in the May~6 data set and
this  prevents accurate
fits to the six lines in this region -- hence the gap in the data
points in the right panel of Fig.~\ref{fig.s11-fe11}.  There are also
isolated pixels where poor fits are found, typically due to missing
pixels or weak lines.

With the intensity of \ion{S}{xi} \lam188.68 measured, one can
estimate the intensity of \ion{S}{xi} \lam186.84 by simply multiplying
by the factor 0.34. This derived intensity can then be compared with the
measured \ion{S}{xi}--\ion{Fe}{xii} feature at 186.88~\AA. For the two
data columns in the May 3 and May 6 datasets analysed in
Sects.~\ref{sect.may6} and \ref{sect.may3}, the results are shown in
Fig.~\ref{fig.s11-fe11}. The key point to note is that the majority of the
pixels show a \ion{S}{xi} contribution of $\le 5$~\%.  Where the
electron density is high in the May 3 data (pixels 50--170, see
Fig.~\ref{fig.fe12-may3}) the \ion{S}{xi} contribution is around 2~\%\
as the \ion{Fe}{xii} lines are more sensitive to higher densities than
the \ion{S}{xi} line.

The contribution of the \ion{S}{xi} line at the 5~\%\ level would have
the most effect on the \ion{Fe}{xii} densities where the \ion{Fe}{xii}
ratio is most sensitive, i.e., around $\log\,N_{\rm e}=10.5$. Using
CHIANTI we find the \ion{Fe}{xii} density is reduced by 0.08~dex. At
$\log\,N_{\rm e}=9.0$ the density is reduced by 0.05~dex. We also note
that \citet{brown08} list a \ion{Ni}{xiv} line as possibly blending
with \ion{S}{xi} \lam188.68, potentially compromising the use of the
\ion{S}{xi} 
line for estimating the \lam186.84 contribution.
Together with the fact that \ion{S}{xi} \lam188.68 is a weak
line that is difficult to measure, we have decided \emph{not} to
correct \ion{Fe}{xii} \lam186.88 for the \ion{S}{xi} contribution in the present
analysis. 

For fitting the 
186.88~\AA\ feature, it is necessary to consider nearby lines  that are not 
directly blended but can affect line fitting. \ion{Fe}{viii}
\lam186.60 is not a blend, but sufficiently close that it can affect
background estimates when fitting \lam186.88. For this work both lines
have been fitted simultaneously with two independent Gaussians. A weak
line is found in the long wavelength wing of \lam186.88
at 186.98~\AA\ which has been suggested to be a \ion{Ni}{xi}
transition \citep{brown08}. Checks on the May 3 and May 6 data sets
have demonstrated 
that this line is always weak relative to \lam186.88, at most around
20~\% of the intensity. Adding an extra Gaussian to the fit to account
for this line often works well, however there are a number of pixels
that yield poor fits if there are missing data or weak lines. For the
present work it was decided to ignore the \ion{Ni}{xi} line by
omitting the pixels at this wavelength -- an example can be seen in
Fig.~\ref{fig.186-example}. This method leads to much more robust fits
while still yielding accurate estimates of the \lam186.88 intensity.

\subsection{Fe\,XII 195.12}\label{sect.195}

The \ion{Fe}{xii} emission line at 195.12~\AA\ is usually the
strongest line observed by EIS and it dominates the spectrum in the
region 194--196~\AA. Despite this, some care has to be taken when
automatically fitting the line. Firstly, there is another
\ion{Fe}{xii} transition at 195.18~\AA\ which lies within the
\lam195.12 profile and thus distorts the observed line profile from a
Gaussian shape. This line is predicted by CHIANTI to be $\le$10~\%\ of
the strength of \lam195.12 below $10^{10}$~cm$^{-3}$, but this
increases to 22~\%\ at $10^{11}$~cm$^{-3}$ and so can not be ignored
if the data set shows high densities. A detailed discussion of the
effects of the blending
\lam195.18 line is provided in Appendix~\ref{app.195}.

\begin{figure}[h]
\centerline{\epsfxsize=9cm\epsfbox{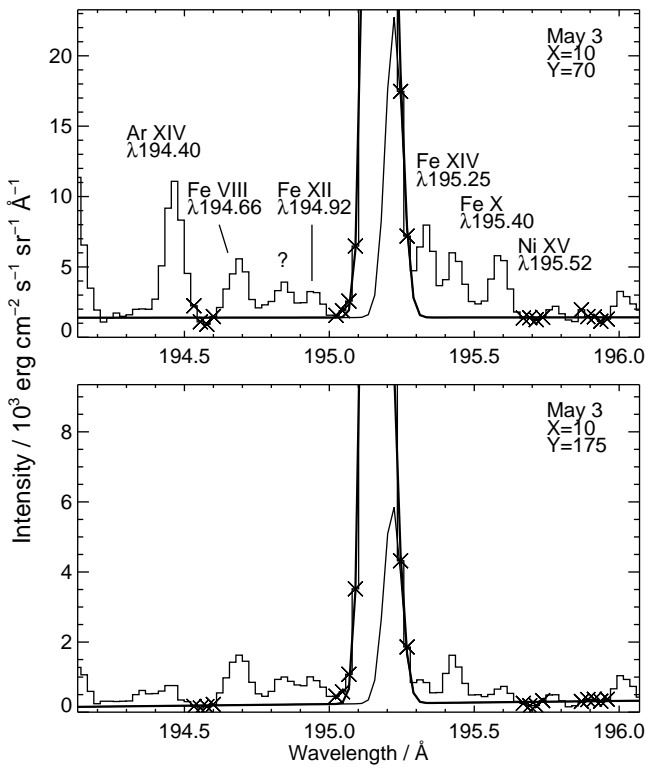}}
\caption{Two example spectra from the May~3 data set showing the
  emission lines neighbouring the strong \ion{Fe}{xii} \lam195.12
  line. A two Gaussian fit has been performed to \lam195.12 and the
  neighbouring \ion{Fe}{xii} \lam195.18 line. The crosses show the
  wavelength pixels that have been included 
  in the fit, and the thick line shows the fit result. A thin line is
  used to highlight the weaker Gaussian that represents \lam195.18.
  Note that there are a number of weak lines neighbouring the
  \ion{Fe}{xii} feature that potentially affect the fits. The peak of
  the \lam195.12 line extends by a factor five beyond the top of each
  of the two plots.}
\label{fig.195-example}
\end{figure}

In addition to the \ion{Fe}{xii} blending component there
are a number of weak lines either side of \lam195.12 in the spectrum,
as can be seen in the two panels in Fig.~\ref{fig.195-example}. Simply
performing a single Gaussian fit over the wavelength range
194.80--195.50~\AA, for example, will lead to an incorrrect fit to the spectrum
background which will result in the intensity of the \lam195.12 line
being underestimated at the 1--3~\%\ level.

The solution to these two problems in the present work is as
follows. Spectral regions are identified that are free from emission
lines and are used to constrain the continuum level. The regions
containing the weak lines are not included in the fit. This is
illustrated in Fig.~\ref{fig.195-example} where crosses indicate the
pixels that have been used for the Gaussian fitting. To deal with the 
\lam195.18 blend a two Gaussian fit has been performed. The stronger
\lam195.12 line has free parameters for the centroid, width and
amplitude, while \lam195.18 is forced to be 0.06~\AA\ to the long
wavelength side of \lam195.12 (the CHIANTI wavelengths for the two
transitions are 195.119 and 195.179~\AA), and to have the same width
as \lam195.12. The amplitude of \lam195.18 is free to vary.

While \lam195.12 is the strongest \ion{Fe}{xii} line observed by EIS
and thus likely to be the most commonly used line for density
diagnostics, there are two other strong \ion{Fe}{xii} lines at
192.39~\AA\ and 193.51~\AA\ that are density insensitive relative to
\lam195.12 and thus can be substituted for this line to be used as
density diagnostics relative to \lam186.88 and \lam196.64. These lines
are unblended and are less likely to saturate on the detector than
\lam195.12 since they are both weaker transitions and the EIS
effective area is lower at these wavelengths. \lam192.39 and
\lam193.51 are discussed in Appendix~\ref{app.fe12} where it is shown
that the ratios of \lam192.39, \lam193.51 and \lam195.12 closely match
the predictions of CHIANTI and thus they all can be used in density
diagnostics.

\subsection{Fe\,XIII 196.54, Fe\,XII 196.64}\label{sect.196}

These two lines are fitted simultaneously with a double Gaussian, with
each parameter free to vary. The main difficulty is to identify a clean
continuum area in the spectrum for accurately fitting the background
for the fits. To the long wavelength side of
\lam196.64 there are three lines: \ion{S}{x} \lam196.81, \ion{Fe}{xii}
\lam196.92  and an unknown line at around 197.02~\AA\
\citep{brown08}. Beyond this 
latter line there is a small region at around 197.2~\AA\ which is free
of lines and suitable for estimating the background level. On the
short wavelength side of \lam196.54 there is a \ion{Fe}{vii} line at
196.42~\AA\ \citep{brown08} but this is generally negligible in active
region conditions.

\begin{figure}[h]
\centerline{\epsfxsize=9cm\epsfbox{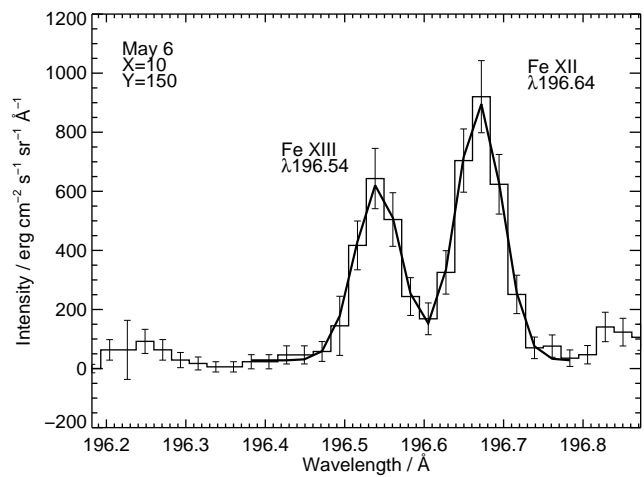}}
\caption{An example double-Gaussian fit to \ion{Fe}{xiii} \lam196.54 and
  \ion{Fe}{xii} \lam196.64  from the May 6 data set. The chosen
  spatial pixel is indicated. The data are
  plotted as a thin line and 
  the fit with a thick line. The error bars on the intensity
  measurements are also shown.}
\label{fig.196-example}
\end{figure}

For the fits performed in this work, only the pixels directly
neighbouring the \ion{Fe}{xii} and \ion{Fe}{xiii} lines were used for
estimating the background level (Fig.~\ref{fig.196-example}) as the
wavelength window was not wide enough to observe the background
region around 197.2~\AA. A uniform background was assumed for the
fitting rather than a linear fit in order to prevent the \ion{S}{x}
\lam196.81 line distorting the derived background.

In regions where there is an absence of hot plasma and the density is
low, the \ion{Fe}{xiii} \lam196.54 line can be very weak and thus
difficult to fit, but this was a not a problem for the data sets
considered here.

\citet{brown08} list a \ion{Fe}{viii} line at 196.65~\AA\ which thus
blends with \ion{Fe}{xii} \lam196.64. CHIANTI predicts the
\ion{Fe}{viii} line to be a factor 0.06 weaker than \ion{Fe}{viii}
\lam194.66 (visible in Fig.~\ref{fig.195-example}), and thus it would
be expected 
to be negligible in most circumstances. However, in active regions
\ion{Fe}{viii} is often seen to be strongly enhanced in small
brightenings or loop footpoints \citep{young07a}, and in these regions
\lam196.65 can be significant. Examples are found in both the
May~3 and May~6 data sets where methods of correcting the
measured \lam196.64 intensity for the \ion{Fe}{viii} contribution are
discussed.

\subsection{Fe\,XIII 202.04}\label{sect.202}

\begin{figure}[h]
\centerline{\epsfxsize=9cm\epsfbox{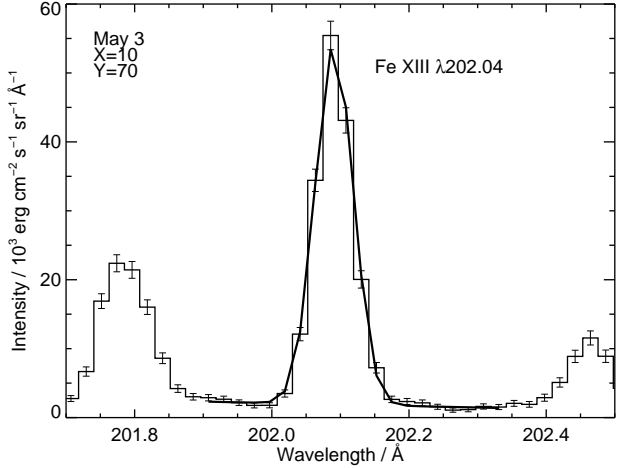}}
\caption{An example single Gaussian fit to the \ion{Fe}{xiii} \lam202.0
emission line. The
May 3 data set is used, and the image pixel indicated.}
\label{fig.202-example}
\end{figure}

\ion{Fe}{xiii} \lam202.04 is unblended and can be fit with a
single Gaussian. We choose to fit a single Gaussian to the
wavelength region 201.90--202.32~\AA\
(Fig.~\ref{fig.202-example}). 

\subsection{Fe\,XIII 203.82}\label{sect.203}

\begin{figure}[h]
\centerline{\epsfxsize=9cm\epsfbox{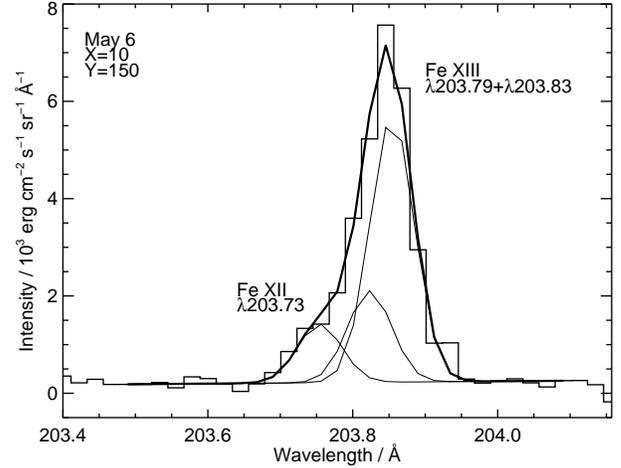}}
\caption{An example fit to the feature at 203.8~\AA\ that comprises 
  \ion{Fe}{xii} \lam203.72, and \ion{Fe}{xiii} \lam\lam203.79, 203.83
  from the May~6 data set. The spatial pixel chosen is
  indicated. Three thin lines are used to show the three individual
  Gaussians from the fit.}
\label{fig.203-example}
\end{figure}

\begin{figure}[h]
\centerline{\epsfxsize=9cm\epsfbox{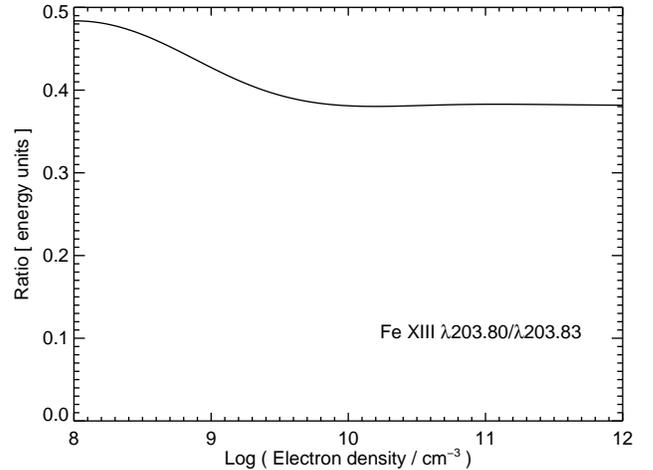}}
\caption{Theoretical variation of the \ion{Fe}{xiii}
  \lam203.80/\lam203.83 ratio derived from CHIANTI.}
\label{fig.fe13-203-lines}
\end{figure}

This feature is a self-blend of two \ion{Fe}{xiii} lines with CHIANTI
wavelengths of 203.797 and 203.828~\AA. The two lines have weak
density dependence (Fig.~\ref{fig.fe13-203-lines}), with the ratio
almost constant above $10^{9.5}$~cm$^{-3}$.
There
is an additional line due to \ion{Fe}{xii} found at 203.728~\AA\ which
is not clearly separated from the \ion{Fe}{xiii} lines
(Fig.~\ref{fig.203-example}). Generally this
line is always weaker than the \ion{Fe}{xiii} feature, but in quiet
Sun conditions it can be comparable in strength.

The fitting method chosen for these lines is to fit the wavelength
region 203.5--204.1~\AA\ with three Gaussians each set to the same
width (which is free to vary). The two \ion{Fe}{xiii} 
components are assumed to have a fixed ratio \lam203.797/\lam203.828
of 0.40. Although Fig.~\ref{fig.fe13-203-lines} demonstrates that this
is only valid for $\log\,N_{\rm e} \gtrsim 9.0$, the assumption is
preferable to allowing the line ratio to freely vary when performing
automatic fits to the data. The close separation of the two
\ion{Fe}{xiii} lines also means that their precise ratio is not
crucial to fitting the spectral feature.
The peak of the \ion{Fe}{xii} component is free to vary in the fit,
but the centroid is fixed relative to the  203.828~\AA\
line using the separation from the CHIANTI database. 
\ion{Fe}{xiii} \lam203.797 also has a fixed centroid relative to the
\lam203.828 line. In all there are four free parameters, with two
additional parameters for a linear fit to the background.

\begin{figure}[h]
\centerline{\epsfxsize=9cm\epsfbox{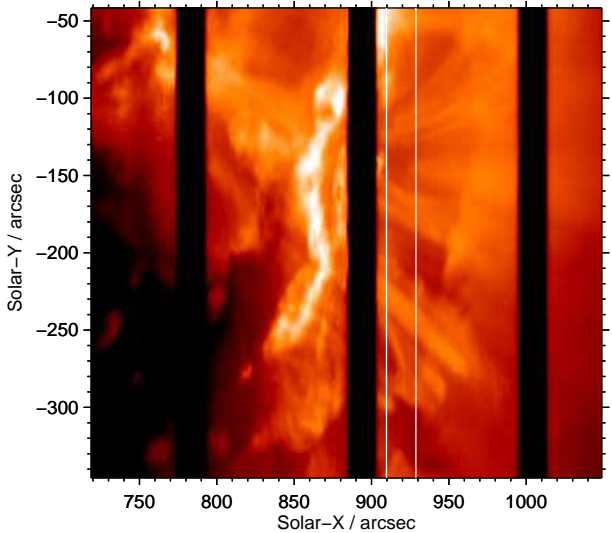}}
\caption{Image of AR 10963 obtained in the \ion{Fe}{xii} \lam195.12
  line on May~6. The raster began at 18:19 UT and completed at
  23:13 and was obtained during the \emph{Hinode} eclipse
  season. The three thick, black, vertical lines show when the
  satellite passed behind the Earth. The two thin vertical black lines
  indicate the spatial region used for the analysis in
  Sect.~\ref{sect.may6}.} 
\label{fig.may6-context}
\end{figure}

\section{Quiescent active region data set (2007 May 6)}\label{sect.may6}

\begin{figure*}[h]
\centerline{\epsfxsize=6.5in\epsfbox{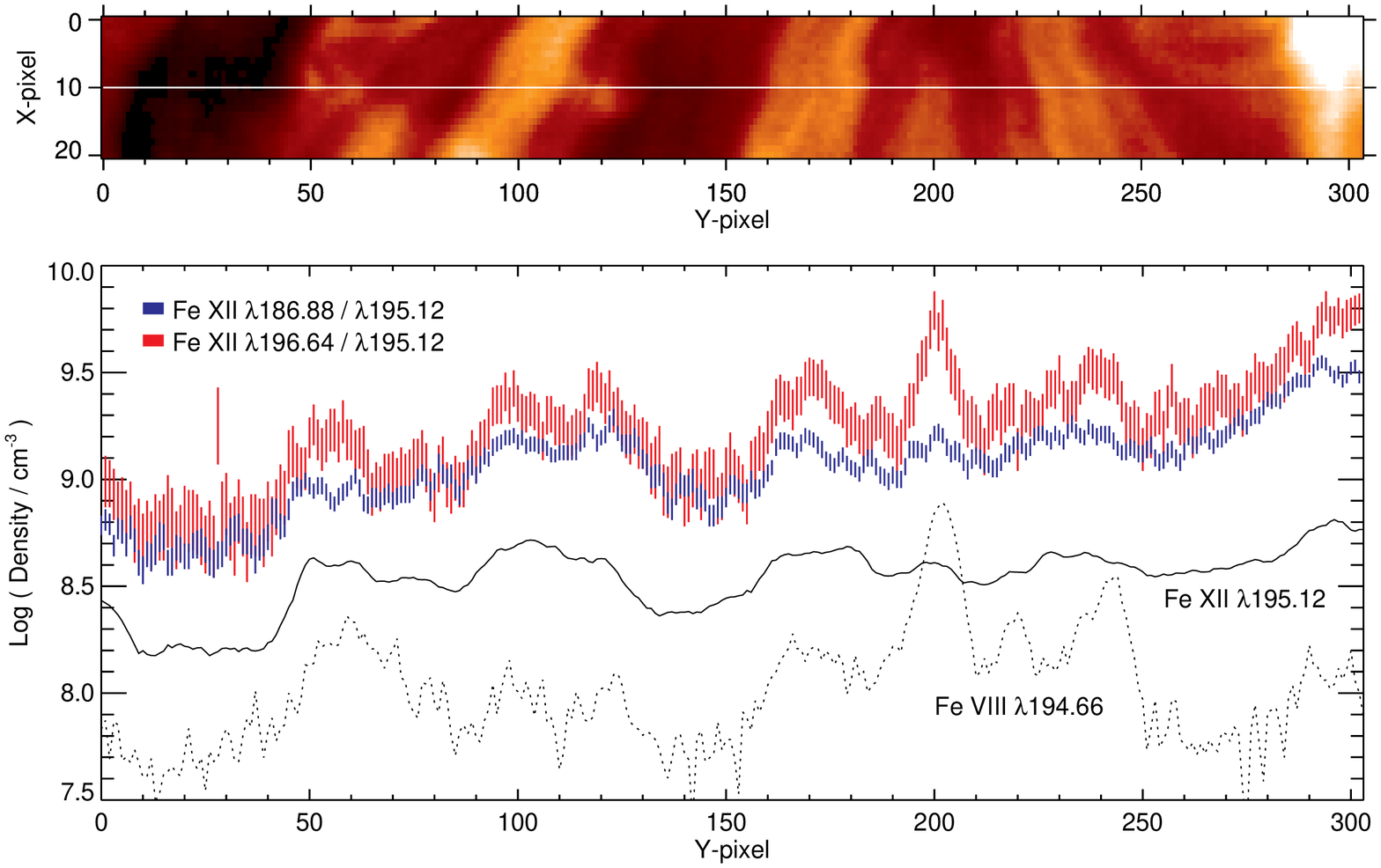}}
\caption{The upper panel shows an image from the May 6 data set
  obtained in the \ion{Fe}{xii} \lam195.12 line. The horizontal line
  indicates the data column chosen to derive densities. The lower panel
  shows densities derived from the \ion{Fe}{xii} \lam186.88/\lam195.12 (blue)
  and \lam196.64/\lam195.12 (red) density diagnostics. Short vertical
  lines are used indicating the lower and upper densities based on the
  1$\sigma$ error bars on the measured line ratios. The black line
  shows the variation of the \lam195.12 line intensity, while the dashed
  line shows the variation of the \ion{Fe}{viii} \lam186.60 line. For the
  latter two lines the quantity plotted is $\log\,(10^{5}I)$ and
  $\log\,(10^{5.5}I)$, respectively, where $I$
  is the line intensity in units \ecss.}
\label{fig.fe12-may6}
\end{figure*}

\begin{figure*}[h]
\centerline{\epsfxsize=6.5in\epsfbox{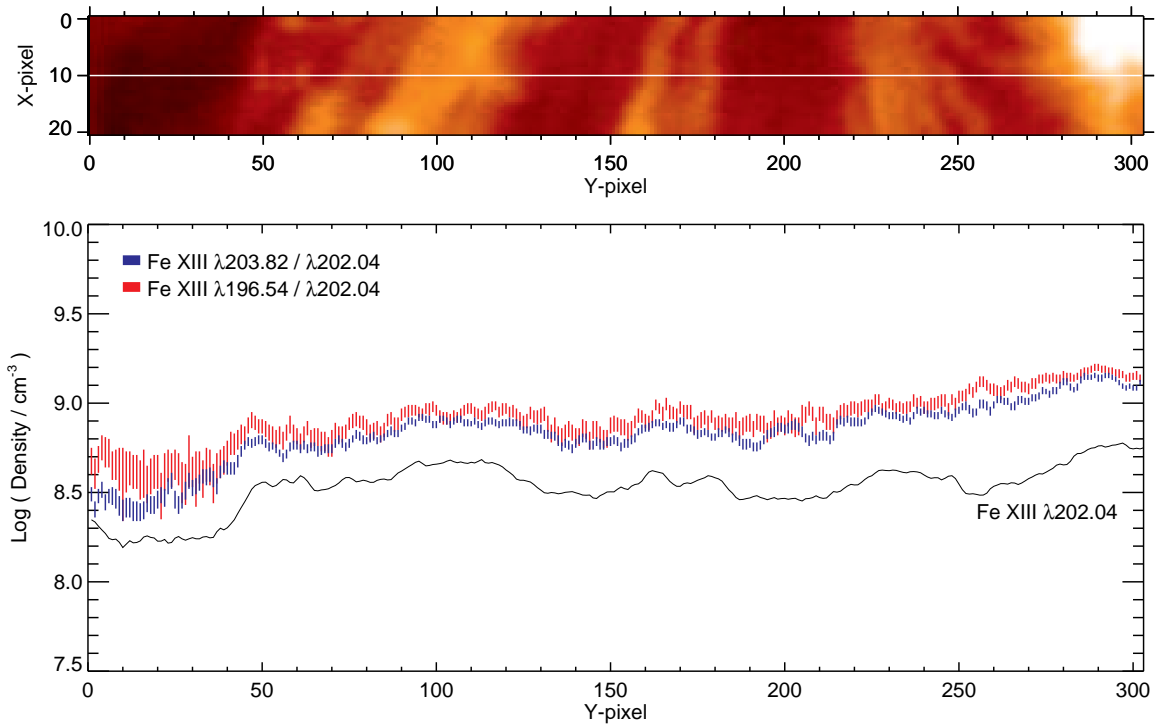}}
\caption{The same plot as Fig.~\ref{fig.fe12-may6} except for
  \ion{Fe}{xiii}. In the upper panel the image shown is from
  \ion{Fe}{xiii} \lam202.02. In the lower panel the red lines show
  densities derived from the \lam196.54/\lam202.02 ratio and the blue
  lines shows densities from the \lam203.82/\lam202.02 ratio. The
  black line shows the variation in intensity of \ion{Fe}{xiii}
  \lam202.04, where the plotted quantity is $\log\,(10^{5}I)$. }
\label{fig.fe13-may6}
\end{figure*}

The data sets presented in this and the next section are from active
region NOAA AR 10963 which rotated across the solar disc during 2007
April 24 to May 9. On May 6 the region was close to the west limb
and observed by EIS with the study AR\_VELOCITY\_MAP which consisted
of an initial context raster with the 40\as\ slit,
followed by a long duration 1\as\ slit raster covering a region
330\as\ $\times$ 304\as. Fig.~\ref{fig.may6-context} shows an image
from the 1\as\ raster formed from \ion{Fe}{xii} \lam195.12. The 
vertical black bands arise from \emph{Hinode} passing behind the Earth
as the raster progressed (the observation was taken during the \emph{Hinode} eclipse season).
Gaussian fits were performed on the spatial region highlighted in 
Fig.~\ref{fig.may6-context}. 
The region has size 21\as\ $\times$ 304\as\ and it crosses
a filament and some coronal loops. At each pixel in this
sub-region each of the density diagnostic lines discussed in
Sect.~\ref{sect.lines} has been fitted. We consider a slice through
the data set at X-pixel 10 and present the derived densities from the
\ion{Fe}{xii} and \ion{Fe}{xiii} lines in Figs.~\ref{fig.fe12-may6}
and \ref{fig.fe13-may6}, respectively. Densities calculated with the
\citet{keenan07} \ion{Fe}{xiii} atomic model are shown in
Fig.~\ref{fig.fe13-may6-agg} (available online). The upper panel in each plot
shows the images derived from the \ion{Fe}{xii} \lam195.12 and
\ion{Fe}{xiii} \lam202.04 lines, respectively, with the selected image
column indicated. 

Considering first the \ion{Fe}{xii} ratios \lam186.88/\lam195.12 and
\lam196.64/\lam195.12, the line intensities from the fitting methods
described in the previous sections are used for each line. \lam186.88
is not corrected for the \ion{S}{xi} \lam186.84 contribution
(Sect.~\ref{sect.186}), but \lam196.64 is corrected for the
\ion{Fe}{viii} \lam196.65 contribution (Sect.~\ref{sect.196}). The
correction was performed by fitting the \ion{Fe}{viii} \lam194.66
emission line at each pixel in the data set. The fitting method for
this line was similar to the nearby \ion{Fe}{xii} \lam195.12
(Sect.~\ref{sect.195}), with only pixel sections chosen for the line
and the background regions at 194.5, 195.7 and 195.9~\AA. From
CHIANTI, the \ion{Fe}{viii} \lam196.65/\lam194.66 ratio is insensitive to
density and has a theoretical value of 0.058. Thus the \ion{Fe}{xii}
\lam196.64 intensity was derived by subtracting 0.058 times the
intensity of \ion{Fe}{viii} \lam194.66 at each pixel. The maximum
contribution of 
\ion{Fe}{viii} to the measured line intensity of the 196.64~\AA\
feature is found to be 9~\% using this method.

The measured intensity ratios were converted to densities using the
CHIANTI database, and are plotted in Figs.~\ref{fig.fe12-may6} and
\ref{fig.fe13-may6}. The 
1$\sigma$ error bars on the intensities have been translated to
errors on the derived densities, and thus vertical lines are used on
these plots, connecting the lower limit to the upper limit.
The 
densities  track each
other reasonably well along the slit, with the positions of peaks and
troughs in good 
agreement. However, the \lam196.64/\lam195.12 ratio generally
gives higher densities than \lam186.88/\lam195.12, and particularly so
at certain of the peaks along the slit, e.g., around pixels 170, 200
and 240. 

The dashed line in Fig.~\ref{fig.fe12-may6} shows the variation of
\ion{Fe}{viii} \lam194.66 along the slit, and it is clear that the
largest discrepancies between the two \ion{Fe}{xii} ratios occur where
\ion{Fe}{viii} \lam194.66 is strongest, implying that the
\ion{Fe}{viii} contribution to \ion{Fe}{xii} \lam196.64 is not fully
accounted for. This suggests that there are problems with the \ion{Fe}{viii}
atomic model in CHIANTI. It is to be noted that adjusting the
theoretical \ion{Fe}{viii} \lam196.65/\lam194.66 ratio to around 0.25
brings the \lam196.64/\lam195.12 and \lam186.88/\lam195.12
densities into good agreement.

The two \ion{Fe}{xiii} density sensitive ratios show excellent
agreement (Fig.~\ref{fig.fe13-may6}), except in the pixel region 0 to
30 where the \lam203.82/\lam202.04 ratio is lower by 0.2--0.3
dex, although the errors bars on the \lam196.54/\lam202.04 ratio are
significantly larger in this region. At all pixels the
\lam196.54/\lam202.04 ratio yields a marginally higher density than
\lam203.82/\lam202.04. Densities derived from the \citet{keenan07}
atomic model (Fig.~\ref{fig.fe13-may6-agg}) are very similar, being
around 0.1~dex lower than for the CHIANTI model, and the two ratios
are in even better agreement, yielding almost identical densities
along the EIS slit.

\begin{figure*}[h]
\centerline{\epsfxsize=6.5in\epsfbox{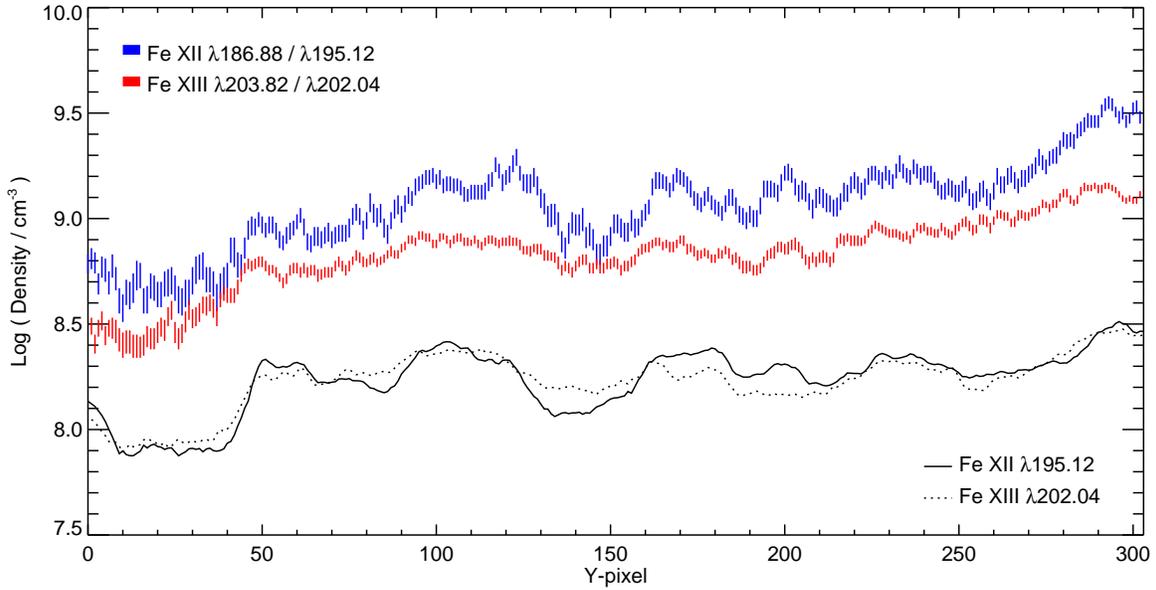}}
\caption{A comparison of densities derived from the \ion{Fe}{xii}
  \lam186.88/\lam195.12 (blue lines) and \ion{Fe}{xiii}
  \lam203.82/\lam202.04 (red lines) ratios. The black solid line shows
  the variation of the \ion{Fe}{xii} \lam195.12 intensity, while the
  black dashed line  shows
  the variation of the \ion{Fe}{xiii} \lam202.04 intensity. For the
  latter two lines the quantity plotted is $\log\,(10^{5.2}I)$, where $I$
  is the line intensity in units \ecss.}
\label{fig.fe12-fe13-may6}
\end{figure*}

In Fig.~\ref{fig.fe12-fe13-may6} we compare the densities from
\ion{Fe}{xii} \lam186.88/\lam195.12 and \ion{Fe}{xiii}
\lam203.82/\lam202.04. (We choose these two ratios for the comparison
due to the \ion{Fe}{viii} correction problem for \lam196.64/\lam195.12
discussed earlier, and the fact that \ion{Fe}{xiii} \lam203.82 is a
stronger line than \lam196.54 at low densities and thus yields
densities with smaller error bars.)
The \ion{Fe}{xii} ratio yields higher densities
at all pixels by up to  0.4 dex, and we also note that
\ion{Fe}{xiii} generally shows less variation pixel-to-pixel than
\ion{Fe}{xii}. These differences will be discussed further in
Sect.~\ref{sect.discussion}.

\onlfig{15}{
\begin{figure*}[h]
\centerline{\epsfxsize=6.5in\epsfbox{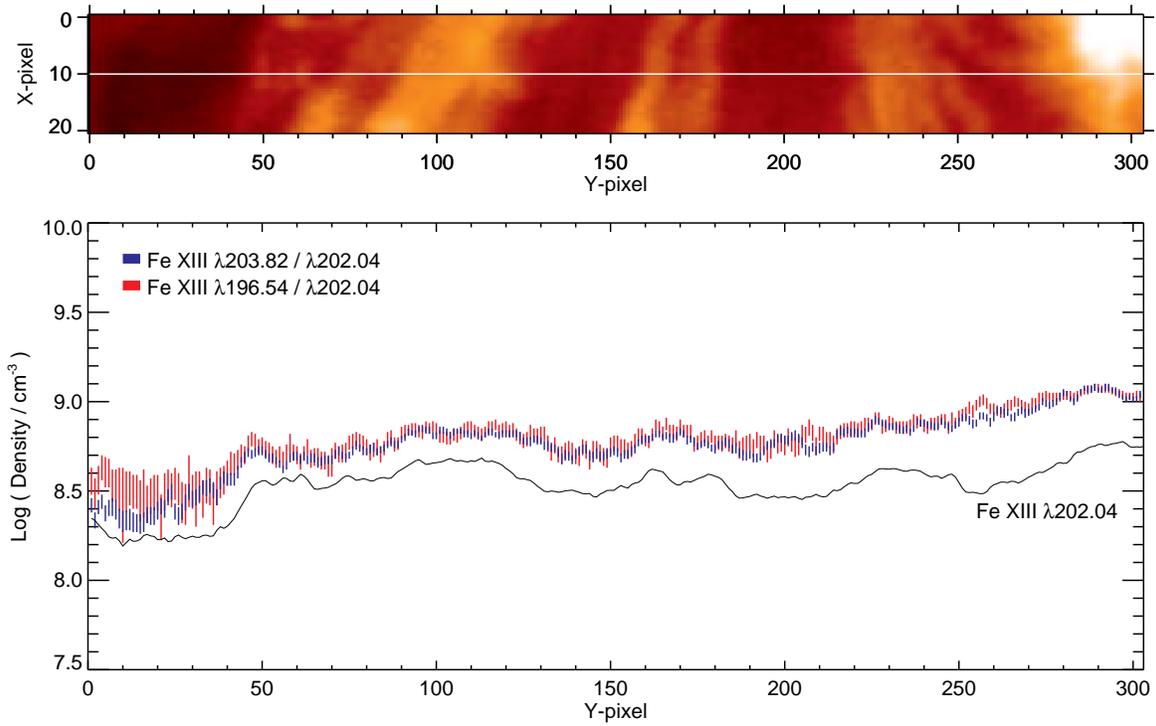}}
\caption{Exactly analogous plot to Fig.~\ref{fig.fe13-may6}, except
  the densities are derived using the \citet{keenan07} \ion{Fe}{xiii}
  atomic model. }
\label{fig.fe13-may6-agg}
\end{figure*}
}

\section{High-density data set (2007~May~3)}\label{sect.may3}

\begin{figure}[h]
\centerline{\epsfxsize=8cm\epsfbox{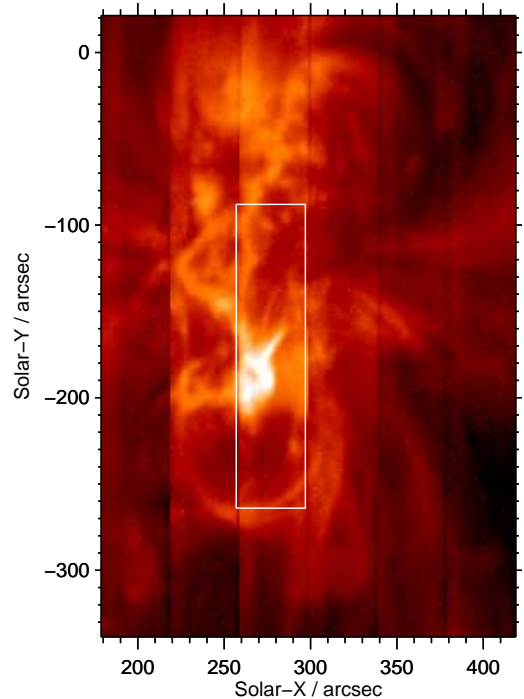}}
\caption{Context image of AR 10963 obtained on 2007 May 3 00:03~UT,
  shortly before the raster analysed in
  Sect.~\ref{sect.may3}. The image was obtained by making a six
  exposure raster with the 40\as\ slit. The image is formed in the
  emission line \ion{Si}{x} \lam261.06 which is formed at a similar
  temperature to \ion{Fe}{xii} \lam195.12. The box indicates the
  location of the 
  narrow slit raster that followed the context raster.}
\label{fig.may3-context}
\end{figure}

\begin{figure*}[h]
\centerline{\epsfxsize=5in\epsfbox{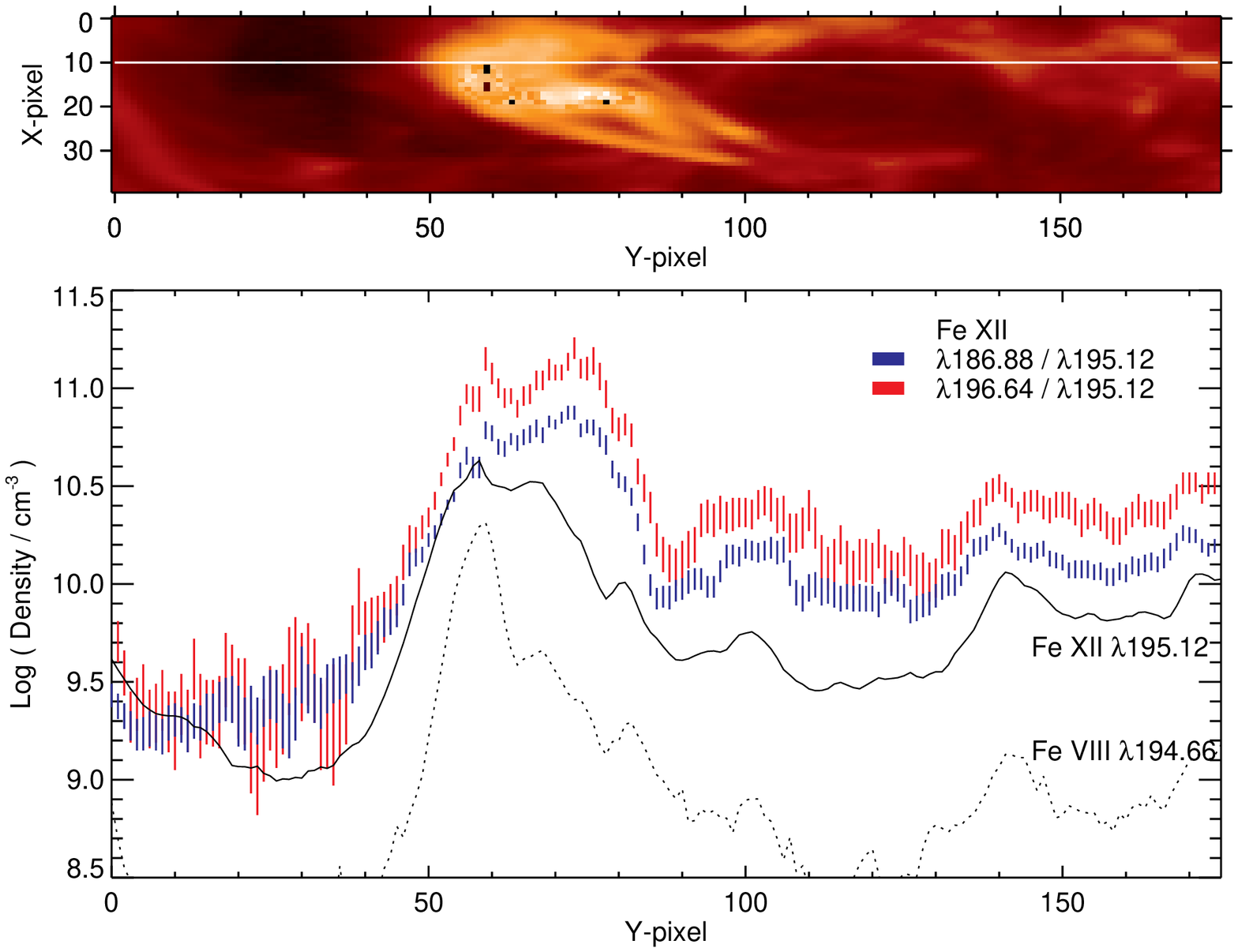}}
\caption{The same plot as Fig.~\ref{fig.fe12-may6} except this time
  for the May 3 data set. For the intensity curves from \ion{Fe}{xii}
  \lam195.12 and \ion{Fe}{viii} \lam194.66, the quantities plotted are
   $\log\,(10^{6.5}I)$ and
  $\log\,(10^{7}I)$, respectively, where $I$
  is the line intensity in units \ecss.}
\label{fig.fe12-may3}
\end{figure*}

\begin{figure*}[h]
\centerline{\epsfxsize=5in\epsfbox{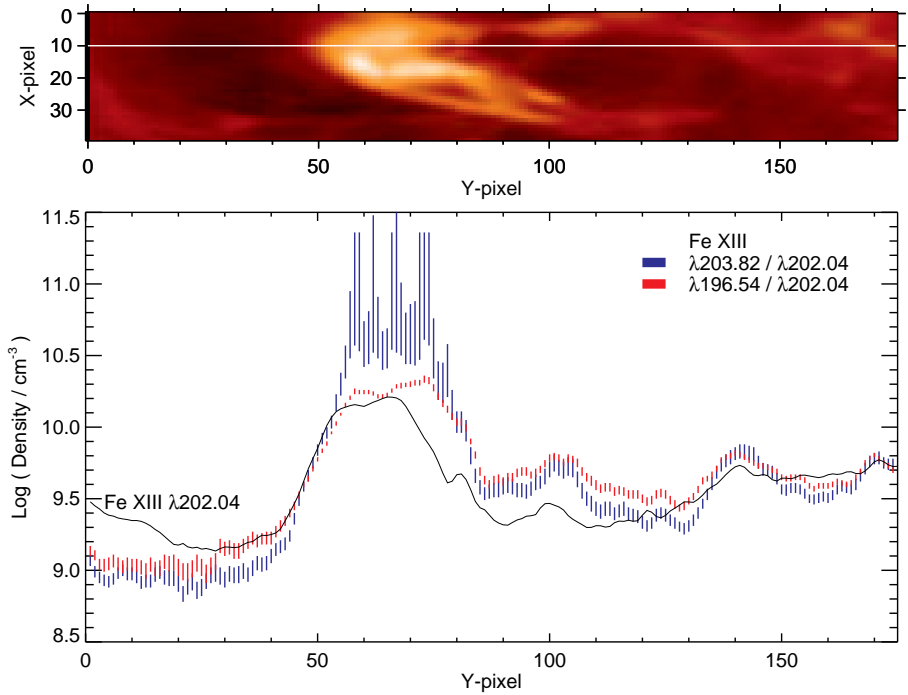}}
\caption{The same plot as Fig.~\ref{fig.fe13-may6} except this time
  for the May 3 data set. For the intensity curve from \ion{Fe}{xiii}
  \lam202.04 the quantity plotted is
   $\log\,(10^{6.5}I)$ where $I$
  is the line intensity in units \ecss.}
\label{fig.fe13-may3}
\end{figure*}

The May~6 data set demonstrates the use of the \ion{Fe}{xii} and
\ion{Fe}{xiii} diagnostics in a quiescent part of an active region,
typical of large, developed coronal loops. The two sets of ion ratios
are sensitive to higher densities, and so the EIS data archive was
searched for examples of high densities. On 2007 May 3, the same
active region showed small flaring activity and this was captured with
the EIS study \diag\ which consists of two rasters, the first
being the same 40\as\ slot context raster used by the AR\_VELOCITY\_MAP
study, and the second being a raster covering an area 40\as\ $\times$
176\as\ using the 
1\as\ slit, the latter having an identical line list to the 
narrow slit raster used on May~6.
\diag\ was run during the period 00:07 to
05:28~UT with the slot context raster run once at the start of the
period, and the narrow slit raster repeated 15 times. A GOES class C9
flare began at May 2 23:15 UT, peaked at 23:50~UT and
decayed by May 3 02:00~UT. The first narrow slit raster ran during the
period 00:11~UT to 00:32~UT during the decay phase of the flare and
the flare line \ion{Ca}{xvii} \lam192.82 is prominent in the
spectrum. Fig.~\ref{fig.may3-context} shows an image obtained with the
context slot raster at 00:07~UT and reveals the large-scale structure of
the active region. The image is from \ion{Si}{x} \lam261.06 as the
\ion{Fe}{xii} and \ion{Fe}{xiii} lines are saturated due to the flare
in this data set.
Saturation of  \lam195.12  also occurs in the brightest parts
of the narrow slit raster image, and so for our density study a data
column is selected that is to one side of the very 
brightest regions in the image
(indicated in the upper panels of Figs.~\ref{fig.fe12-may3} and 
\ref{fig.fe13-may3}). Even in this data column, one or two pixels in
the centre of the \lam195.12 profile are saturated at the
brightest part of the image, over Y-pixels 54 to 69. Surprisingly,
though, the two Gaussian fit to the 195~\AA\ feature still yielded
good results. This was checked in two ways. Firstly the insensitive
\lam192.39/\lam195.12 ratio was plotted along the data column
(Appendix~\ref{app.fe12}) and no significant feature is found over the
saturated region. Secondly, the density was re-computed using the
\lam186.88/\lam192.39 ratio, and differences of at most 0.15~dex were
found.

As for the May~6 data set, \ion{Fe}{xii} \lam196.64 has been corrected
for the \ion{Fe}{viii} blend using the \lam194.66 line, although the
\ion{Fe}{viii} contribution to the \lam196.64 feature is at most 5~\%\
here. 

X-pixel 10 in the May~3 data-set was selected for the density
analysis, and the densities from the pairs of \ion{Fe}{xii} and \ion{Fe}{xiii}
ratios are presented in Figs.~\ref{fig.fe12-may3} and
\ref{fig.fe13-may3}. The \ion{Fe}{xiii} densities derived from the
\citet{keenan07} atomic model are presented in
Fig.~\ref{fig.fe13-may3-agg} (available online).

The derived \ion{Fe}{xii} densities again track each other reasonably
well, with 
 \lam196.64/\lam195.12 yielding higher densities by up to 0.4
dex, with the largest discrepancies being at high densities
(Fig.~\ref{fig.fe12-may3}). A strong peak in the \ion{Fe}{viii}
\lam194.66 line is seen around Y-pixel 57 which may partly account for the
density discrepancy here, but at other locations this can not be the case.

For \ion{Fe}{xiii} the two ratios again track each other very well at
low densities (Fig.~\ref{fig.fe13-may3}), but at high densities \lam203.83/\lam202.04 is
found to get close to the high-density limit of $\log\,N_{\rm e}=11.3$
of the ratio (indicated by the large
error bars). This does not happen for \lam196.54/\lam202.04 which has
greater sensitivity to high densities. This issue will be discussed
further in Sect.~\ref{sect.discussion}. The same problem occurs when
using the \citet{keenan07} atomic model (Fig.~\ref{fig.fe13-may3-agg}), although as the
\lam203.82/\lam202.04 high-density limit is slightly larger for this
model (Fig.~\ref{fig.aggarwal}) then the discrepancy between the two
ratios is not as large at high densities.  Densities from
\lam196.54/\lam202.04 are 0.2~dex lower at the highest densities using
the \citet{keenan07} model compared to the CHIANTI model.

\begin{figure*}[h]
\centerline{\epsfxsize=5in\epsfbox{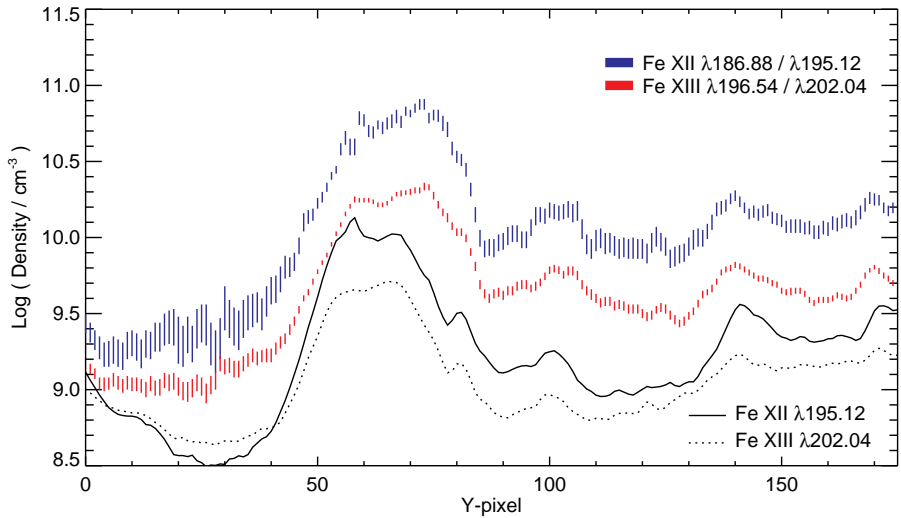}}
\caption{A comparison of densities derived from the \ion{Fe}{xii}
  \lam186.88/\lam195.12 (blue lines) and \ion{Fe}{xiii}
  \lam196.64/\lam203.82 (red lines) ratios in the May 3 data set. The
  intensity curves from \ion{Fe}{xii} \lam195.12 and \ion{Fe}{xiii}
  \lam202.04 show the quantity $\log\,(10^{6}I)$ where $I$
  is the line intensity in units \ecss.}
\label{fig.fe12-fe13-may3}
\end{figure*}

In Fig.~\ref{fig.fe12-fe13-may3} the densities from \ion{Fe}{xii}
\lam186.88/\lam195.12 and \ion{Fe}{xiii} \lam196.54/\lam202.04 are
compared.  \lam196.54/\lam202.04 is preferred to \lam203.82/\lam202.04
for \ion{Fe}{xiii} as we believe \lam203.82/\lam202.04 is not
accurately measuring the density for $\log\,N_{\rm e}\ge 10.0$ (Sect.~\ref{sect.discussion}).
The discrepancies noted for the May~6 data set are also seen
here, with \ion{Fe}{xii} yielding higher densities by up to
0.5~dex. If the \citet{keenan07} model is used the difference between
\ion{Fe}{xii} and \ion{Fe}{xiii} rises to 0.7~dex. The magnitude of
this difference will be discussed in Sect.~\ref{sect.discussion}.

\onlfig{20}{
\begin{figure*}[h]
\centerline{\epsfxsize=5in\epsfbox{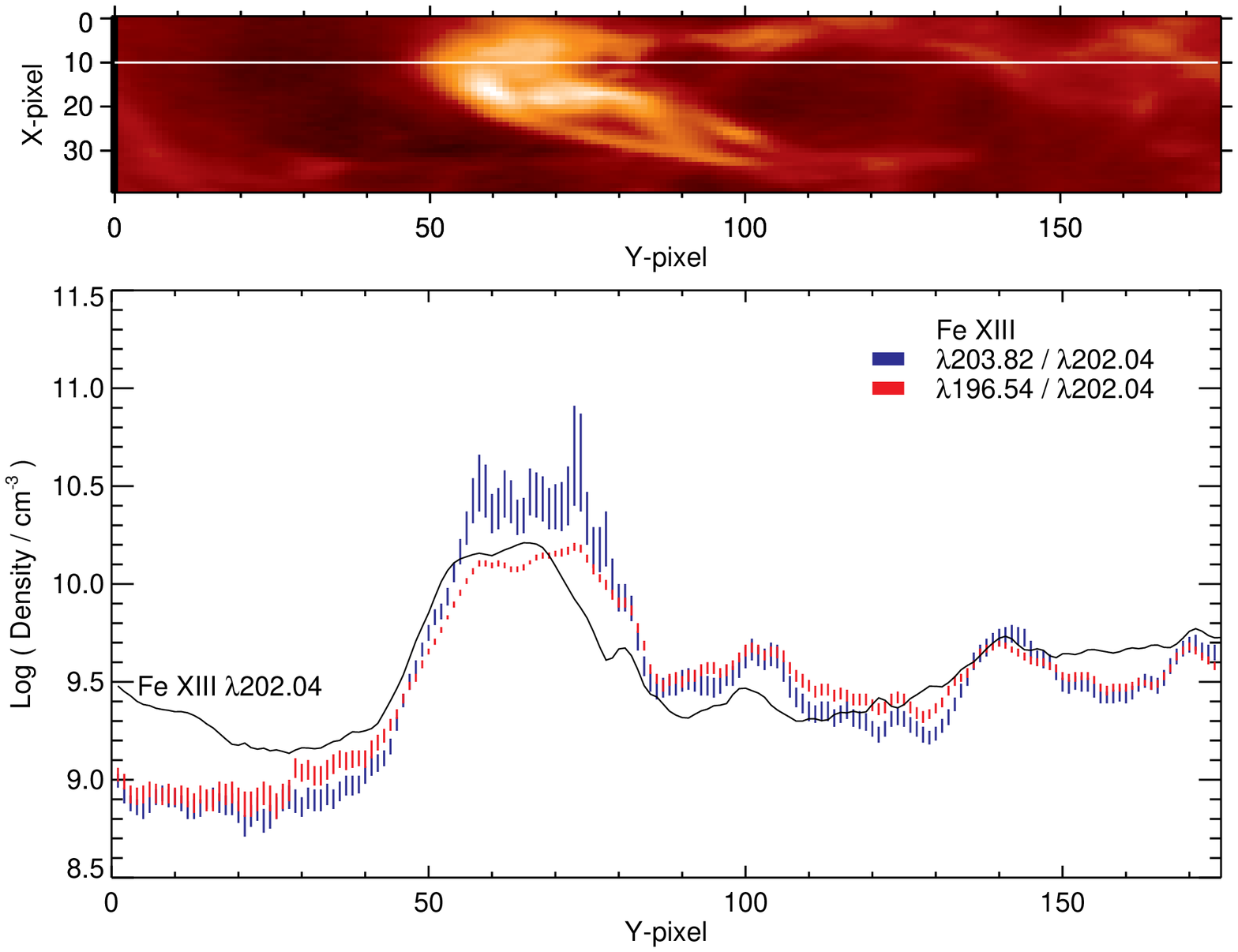}}
\caption{Exactly analogous plot to Fig.~\ref{fig.fe13-may3}, except
  the \ion{Fe}{xiii} densities are derived using the \citet{keenan07}
  atomic model.}
\label{fig.fe13-may3-agg}
\end{figure*}
}

\section{Integrity of the line fit parameters}\label{sect.integrity}

\begin{figure*}[h]
\epsfxsize=7.5in\epsfbox{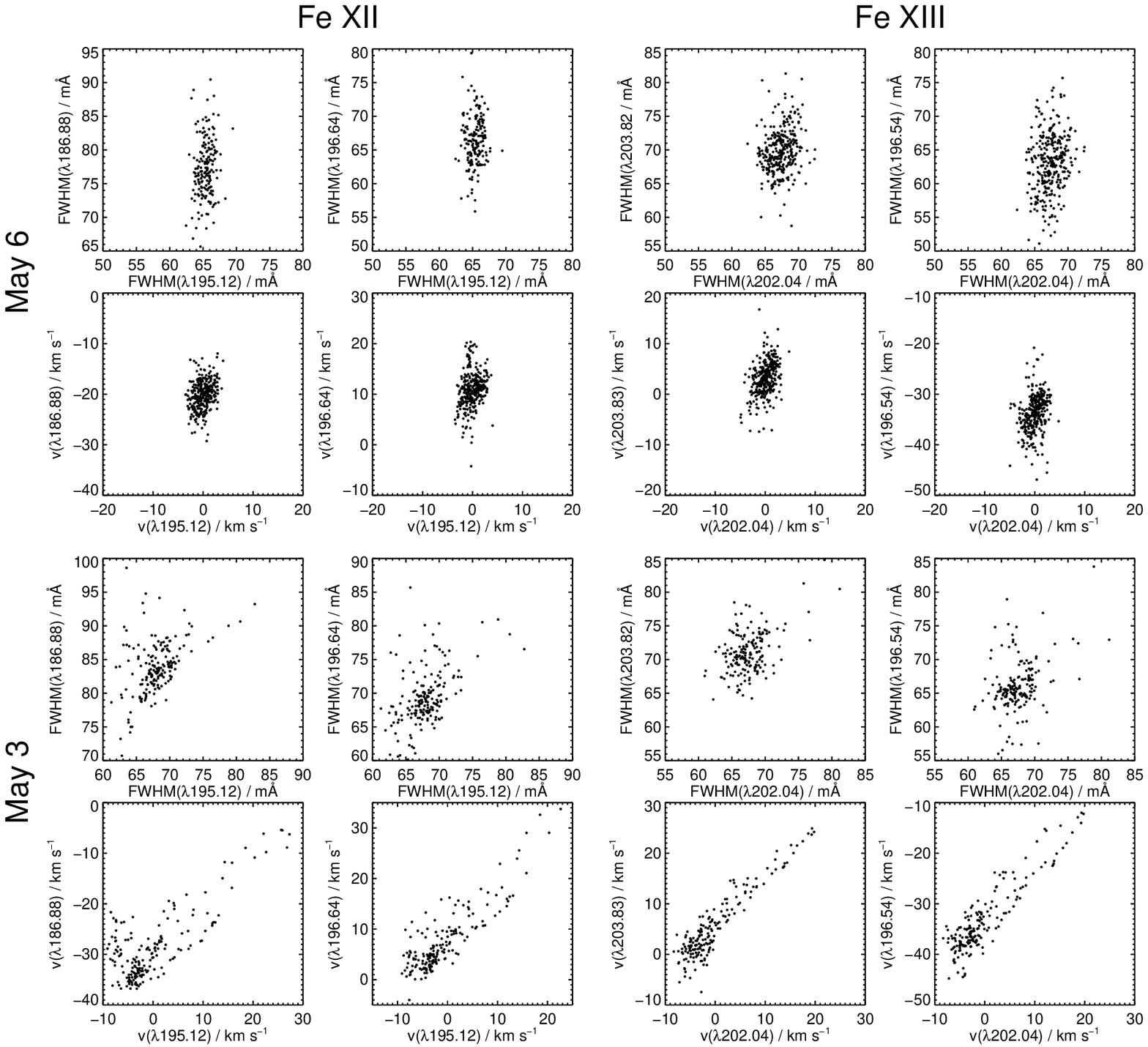}
\caption{Comparisons of line widths and velocities for the density
  diagnostic lines of \ion{Fe}{xii} and \ion{Fe}{xiii} from the May~6
  and May~3 data sets. For \ion{Fe}{xii}, \lam195.12 is used as the
  reference line, while for \ion{Fe}{xiii} \lam202.04 is used. In each
  plot a cross indicates the value of FWHM or velocity from the two
  lines being compared at a particular spatial pixel along the data
  columns discussed in Sects.~\ref{sect.may6} and \ref{sect.may3}. The
  top two rows of plots show comparisons from May~6, while the bottom
  two rows show comparisons from May~3. The two left-most columns
  contain the comparisons for the \ion{Fe}{xii} lines, and the two
  right-most columns the comparison for the \ion{Fe}{xiii} lines.}
\label{fig.fit-comparisons}
\end{figure*}

The density analysis presented in the previous sections used the
emission line intensities that are derived from Gaussian fits to the
emission lines. The fits also yield centroid and line width
information for each line and these are valuable for assessing the
quality of the line fits and for revealing any unaccounted for line
blending. Each of the three \ion{Fe}{xii} lines and each of the three
\ion{Fe}{xiii} lines would be expected to show the same line width and
velocity behaviour across the data and the plots in
Fig.~\ref{fig.fit-comparisons} show comparisons for the various line
pairs. 
For \ion{Fe}{xii},
the comparisons are done against \lam195.12 (the eight panels on the
left side of Fig.~\ref{fig.fit-comparisons}), while for \ion{Fe}{xiii}
they are done against \lam202.04 (the eight panels on the
right side of Fig.~\ref{fig.fit-comparisons}). It is to be noted that
the measured \lam186.88 line is a blend of two \ion{Fe}{xii} emission
lines separated by 0.033~\AA\ which thus affects the width and
velocity measurements. \ion{Fe}{xiii} \lam203.82 was assumed to be
a blend of two \ion{Fe}{xiii} lines when calculating densities,
however, in the Gaussian fitting (Sect.~\ref{sect.203}) the line
properties of both components were derived thus the comparisons in
Fig.~\ref{fig.fit-comparisons} are for the longer wavelength
\lam203.828 component.

Considering first the line widths (the full width at half maximum,
FWHM, in this case) we plot the widths of the line of interest against
the reference line in row 1 and row 3 of
Fig.~\ref{fig.fit-comparisons} for the May~6 and May~3 data sets,
respectively. Ideally the points should lie on a diagonal line running
from the bottom-left to top-right, indicating that all the
\ion{Fe}{xii} or \ion{Fe}{xiii} lines broaden together. For the May~6
data set the line width of \lam195.12 is very consistent along the
slit, with an average width of $66.0\pm 1.3$~m\AA.
However,
both \lam186.88 and \lam196.64 show a significantly broader spread,
with average values of $78.0\pm 4.1$~m\AA\ and $67.3\pm 4.5$~m\AA. The
greater spread is simply due to the lower signal-to-noise of the
two lines. The \lam196.64 and \lam195.12 widths are in excellent
agreement, while the larger width of \lam186.88 arises from the fact
that it is a blend of two lines with wavelengths 186.85 and
186.89~\AA. 

For \ion{Fe}{xiii}, the greater spread of widths for \lam203.83 and
\lam196.54 are again explained by the lower signal-to-noise in these
lines. The average width for \lam202.04 is $67.4\pm 1.7$~m\AA, while
for \lam203.83 and \lam196.54 they are $69.9\pm 3.4$~m\AA\ and $63.2\pm
5.0$~m\AA, respectively. The latter is noticeably lower than the other
width measurements, although it is consistent within the error bars.

Moving to the May~3 data set, a greater spread of line widths is found
particularly for \ion{Fe}{xii}. The average widths for \lam195.12,
\lam186.88 and \lam196.64 are $67.9\pm 3.1$, $84.0\pm 4.4$ and
$69.4\pm 4.2$~m\AA, respectively. The average \lam186.88 width is significantly
larger than for  May~6 due to an increasing contribution of the
186.85~\AA\ component at high densities. The other two lines have
comparable widths to May~6, only slightly broader as might be expected
for this flare data set.  The average widths for \ion{Fe}{xiii}
\lam202.04, \lam196.54 and \lam203.83 are $67.5\pm 2.8$, $66.0\pm 3.9$
and $71.0\pm 3.1$~m\AA, respectively. We note that, as with the May~6
data set, \lam203.82 is a little broader than the other lines which
implies the three Gaussian model for the lines at this wavelength may
not be correct. Either an unaccounted for blending line, or errors in
one or more of the rest wavelengths could be responsible. The effect,
however, is small.

For comparing the line centroid measurements we consider the line
velocities as given by
\begin{equation}
v=c { \lambda-\lambda_{\rm ref} \over \lambda_{\rm ref}}
\end{equation}
where $c$ is the speed of light, $\lambda$ is the measured emission
line centroid and
$\lambda_{\rm ref}$ is the reference wavelength of the line (taken
from CHIANTI). The positions of the EIS emission lines on the detector
vary over the orbit by $\approx \pm 1$~pixels \citep{brown07}, and
so the measured centroids are normalised here relative to the
reference lines (\lam195.12 for \ion{Fe}{xii} and
\lam202.04 for \ion{Fe}{xiii}), i.e., the wavelengths of these lines
along the EIS slit
are assumed to average to the reference wavelengths of the lines. As
for the line widths, the line velocities should ideally fall on a diagonal
line running from bottom-left to top-right in the figures, indicating
that each line should reveal the same blueshift or redshift. In
addition the diagonal line should pass through (0,0) if the reference
wavelengths of the lines are both correct.

For the May~6 data set there is little dynamic activity along the slit
which is revealed in the \lam202.04 and \lam195.12 measurements which
each have a very narrow spread of values. The
other lines show a broader spread of values due to the lower
signal-to-noise of these lines. \ion{Fe}{xii} \lam196.64 
shows a small clump of values around $+20$~\kms\ which is caused by
the blending \ion{Fe}{viii} line mentioned in
Sect.~\ref{sect.196}. Of the four comparison plots for May~6, only for
\lam203.83 are the points close to the (0,0) point, indicating
accurate reference wavelengths. For \lam186.88 the $\approx -20$~\kms\
offset is explained by the blending 186.85~\AA\ line. For \lam196.64
the average offset is $10.9\pm 3.4$~\kms\ which suggests a revised
reference wavelength for the line of 196.647~\AA\ (assuming the
CHIANTI wavelength for \lam195.12 is correct). This compares with the
value of 196.645~\AA\ in the quiet Sun spectrum of \citet{brown08}. The average offset for
\ion{Fe}{xiii} \lam196.54 is $-33.9\pm 3.9$~\kms, suggesting a revised
reference wavelength of 196.518~\AA\ (assuming the
CHIANTI wavelength for \lam202.04 is correct). The average value in
the five spectra presented by \citet{brown08} is 196.520~\AA. The
uncertainties in the reference wavelengths of \ion{Fe}{xii} \lam195.12
and \ion{Fe}{xiii} \lam202.04 are $\pm 0.002$~\AA\
\citep{delzanna05,brown07}, thus our revised wavelengths for the
\ion{Fe}{xii} \lam196.64 and \ion{Fe}{xiii} \lam196.54 are $196.647\pm
0.003$~\AA\ and $196.518\pm 0.003$~\AA, respectively.

The line velocity plots from the May~3 data set in the bottom row of
Fig.~\ref{fig.fit-comparisons} are quite striking. There is clearly
significant dynamic activity in the data set which is actually related
to the high density regions around Y-pixels 40 to 90. Reassuringly a
clear bottom-left to top-right diagonal is seen in each plot,
demonstrating that the \ion{Fe}{xii} and \ion{Fe}{xiii} lines are
responding to the velocity shifts in unison. The effect is
particularly clear for \ion{Fe}{xiii} \lam203.83 and \lam202.04. In
addition to the diagonal pattern, it is also clear that the velocity
offsets discussed above for the May~6 data set are consistent with
those for May~3, confirming the integrity of the EIS wavelength scale
over time.  An interesting effect in the velocity plots for May~3 is
the loop-like structure visible for the red-shifted pixels and most clearly
seen in the \ion{Fe}{xiii} \lam196.54--\lam202.04 plot, but also
partly seen in each of the other three plots. No explanation can be
offered at this point for such an effect.

In summary, the plots shown in Fig.~\ref{fig.fit-comparisons} give
confidence in the quality of the Gaussian fits employed in the
previous sections. In particular we see no evidence of line blending
(beyond that already accounted for) that would lead to anomalous
broadening or line shifts.

\section{Discussion}\label{sect.discussion}

Sects.~\ref{sect.may6} and \ref{sect.may3} compared the derived
densities from the four \ion{Fe}{xii} and \ion{Fe}{xiii} line ratios and
significant 
differences were seen between the two ions. In this section we
consider whether these differences demonstrate real physical
differences between the \ion{Fe}{xii} and \ion{Fe}{xiii} emitting
regions, or whether atomic or instrumental effects could be
responsible. Firstly we consider the pairs of ratios from each species.

\begin{figure}[h]
\centerline{\epsfxsize=9cm\epsfbox{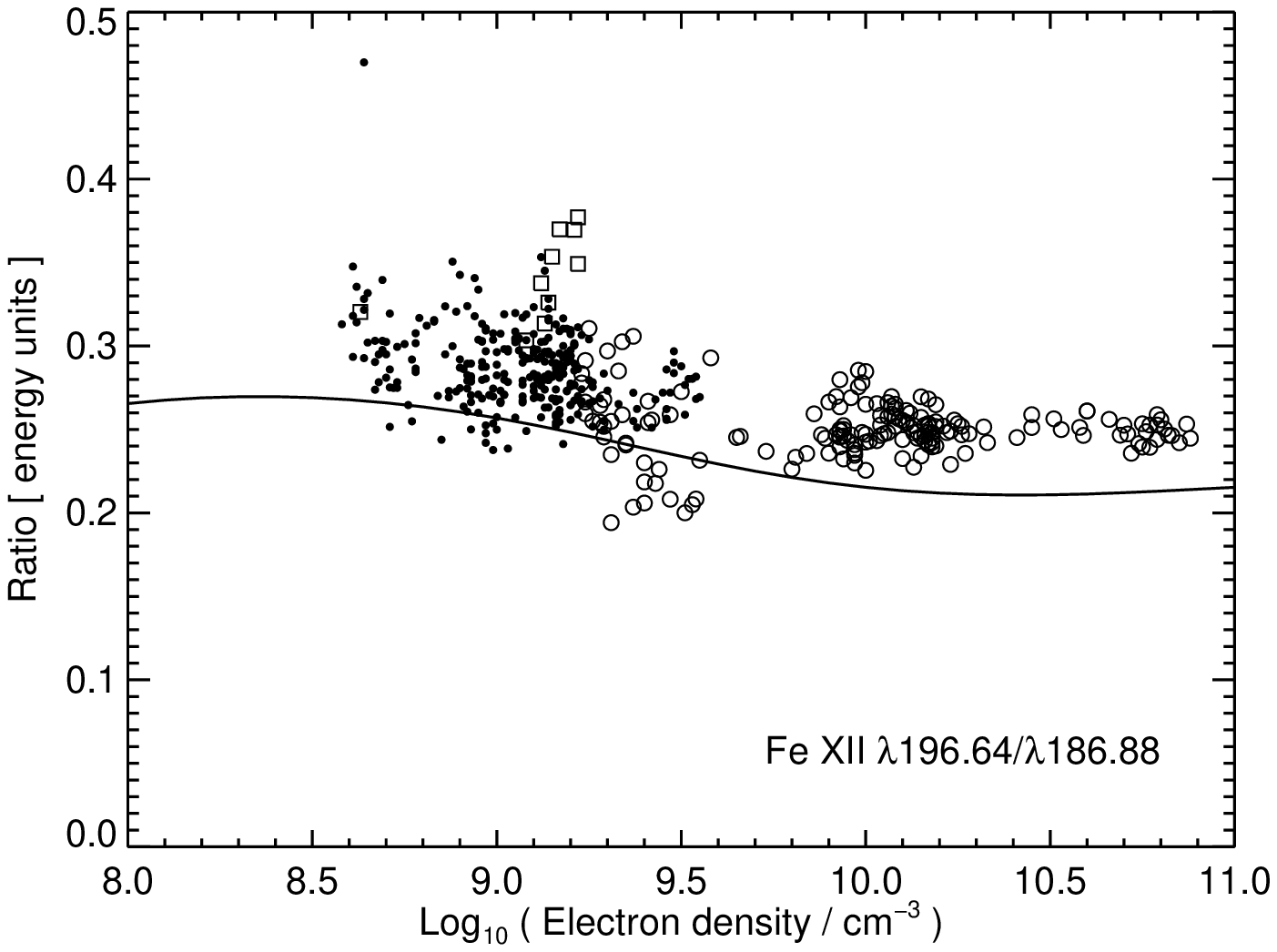}}
\caption{A comparison of the measured variation of the \ion{Fe}{xii}
  \lam196.64/\lam186.88 intensity ratio with density 
  compared to the predictions from the
  CHIANTI database. The measured densities are obtained from the
  \ion{Fe}{xii} \lam186.88/\lam195.12 ratio. Small filled circles denote
  measurements from the May~6 data set, larger open circles denote measurements
  from the May~3 data set. Points from the May~6 data set where the
  \ion{Fe}{viii} is predicted to contribute 6~\%\ or more to the
  blended \lam196.64 line are denoted by open squares.}
\label{fig.196-186}
\end{figure}

\ion{Fe}{xii} \lam196.64/\lam195.12 systematically yields higher
densities than \lam186.88/\lam195.12 in both
data sets. Sect.~\ref{sect.may6} identified a blending \ion{Fe}{viii}
line as being partly responsible, but this does not explain the
consistent discrepancies seen in the May~3 data set. The
\lam196.64/\lam186.88 ratio is actually relatively \emph{insensitive}
to density
(a fact used in Appendix~\ref{sect.offsets} to help estimate the grating
tilt), and Fig.~\ref{fig.196-186} compares how the measured ratio
values from the two data sets compare with the predictions from
CHIANTI. The measured ratio values are plotted against the density
derived from \lam186.88/\lam195.12, with the predicted ratio variation
with density from CHIANTI over-plotted.
It is clearly seen that CHIANTI systematically
under-estimates the measured ratios except for a small group of points
in the May~3 data set. The shape of the gently sloping curve is
reproduced quite 
well, except the high density points are a little further from the
curve, explaining why the difference in derived densities is larger
when the densities are high (Fig.~\ref{fig.fe12-may3}).
The effect of the blending \ion{Fe}{viii} line on \lam196.64 is
clearly seen through the position of the small group of squares in the
plot which denote spatial locations where \ion{Fe}{viii} is estimated
to contribute more than 6~\%\ of the intensity of the measured line at
196.64~\AA. An additional blending line could explain the systematic
offset between the measured and 
predicted values of the \lam196.64/\lam186.88 curve, but such a line
would have to show very similar temperature and density behaviour to
\lam196.64 and this is highly unlikely.

Based on the fact that  \lam186.88/\lam195.12 gives densities closer
to \ion{Fe}{xiii} than \lam196.64/\lam195.12 we believe that the
former ratio is thus more accurate. The discrepancy between
observations and theory  revealed in Fig.~\ref{fig.196-186} is then
due to  CHIANTI under-estimating the \lam196.64/\lam195.12 ratio,
however, a systematic  error in the relative calibration of EIS of around
10--15~\%\ between
wavelengths 186.88~\AA\ and 196.64~\AA\ can not be ruled out.

\begin{figure}[h]
\centerline{\epsfxsize=9cm\epsfbox{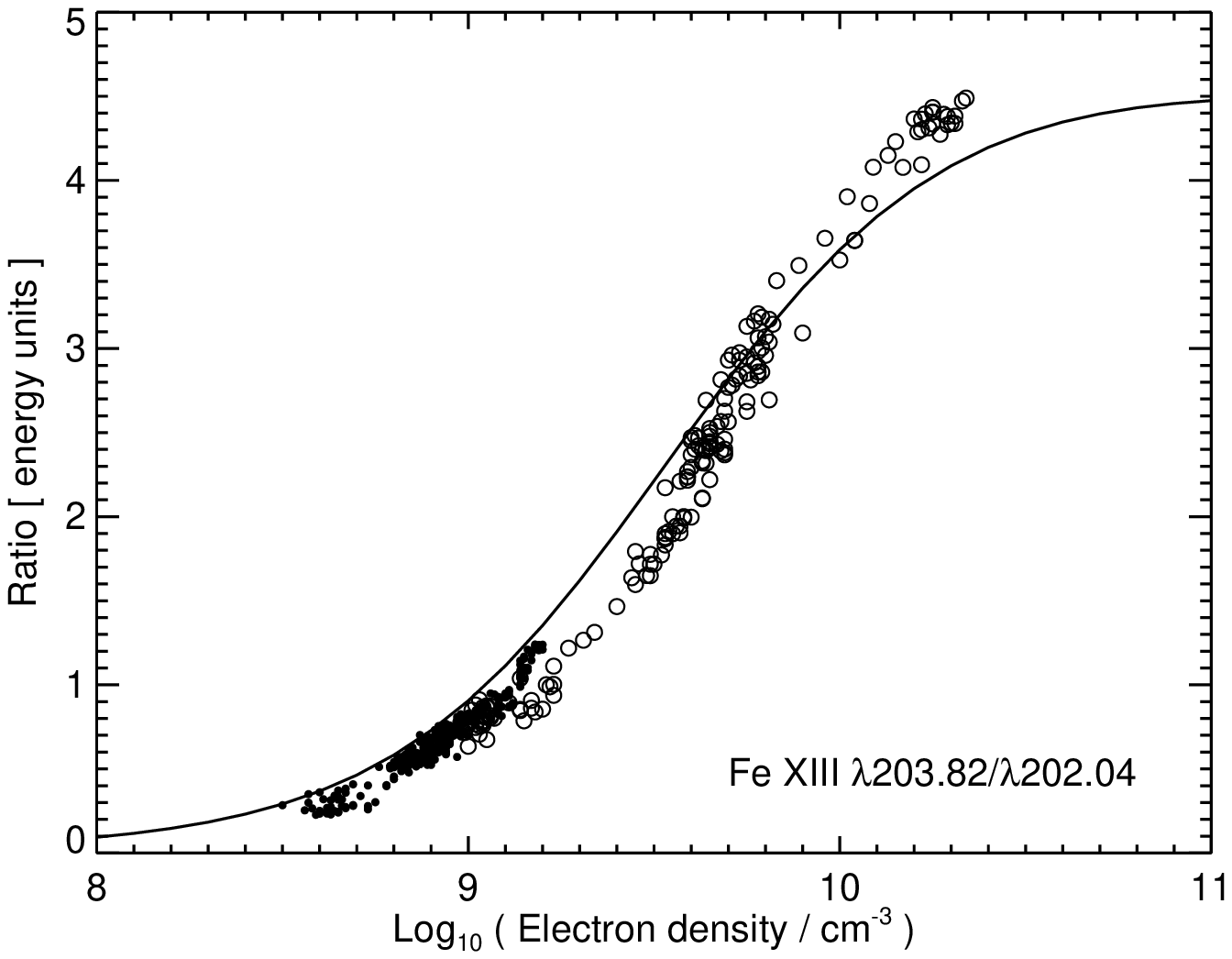}}
\caption{A comparison of the measured variation of the \ion{Fe}{xiii}
  \lam203.82/\lam202.04 intensity ratio with density 
  compared to the predictions from the
  CHIANTI database. The measured densities are obtained from the
  \ion{Fe}{xiii} \lam196.54/\lam202.04 ratio. Small filled circles denote
  measurements from the May~6 data set, larger open circles denote measurements
  from the May~3 data set. }
\label{fig.203-density}
\end{figure}

The discrepancy between the densities predicted by \ion{Fe}{xiii}
\lam203.82/\lam202.04 and \lam196.54/\lam202.04 only becomes
significant at high densities and this clearly  lies in a problem
with the \lam203.82/\lam202.04 ratio, our reason being that the
density variation from \lam196.54/\lam202.04 shown in
Fig.~\ref{fig.fe12-fe13-may3} tracks the variation from \ion{Fe}{xii}
\lam186.88/\lam195.12 very well. The densities from
\lam203.82/\lam202.04 increase anomalously fast compared to the three
other ratios when the density is above $10^{10}$~cm$^{-3}$. We can
investigate this effect a little further by plotting the measured
\lam203.82/\lam202.04 ratio against the densities derived from
\lam196.54/\lam202.04, and comparing with the predicted ratio
variation from CHIANTI (Fig.~\ref{fig.203-density}). 
It is seen that above $10^{10}$~cm$^{-3}$  the
measured ratio values are not actually far from the theoretical curve,
but because the curve flattens at these densities the measured ratios
translate to high densities using the CHIANTI model. A discrepancy of
a similar magnitude is seen around ratio values of 1.5, but this
translates to only a small difference in density due to the high
gradient of the theoretical curve at this location.  A simple remedy
to the high density  problem is if the high-density limit predicted by CHIANTI is
increased from the current value of 4.5 to around 5.0. The
\citet{keenan07} \ion{Fe}{xiii} model predicts a high-density limit of
4.7 (Fig.~\ref{fig.aggarwal}) and so the densities derived from the
May~3 data set are correspondingly lower in the high density region
(Fig.~\ref{fig.fe13-may3-agg}). The  high-density limit
for \lam203.82/\lam202.04 will be a valuable test for future atomic
calculations of \ion{Fe}{xiii}.

Figs.~\ref{fig.fe12-fe13-may6} and \ref{fig.fe12-fe13-may3} showed the
density comparisons between \ion{Fe}{xii} and \ion{Fe}{xiii} for what we
consider to be the best diagnostics for each ion. The features in the
density profiles generally match very well as might be expected for
two ions formed at very similar temperatures, however the absolute
density values differ, with \ion{Fe}{xii} almost always giving higher
densities than \ion{Fe}{xiii}. Also striking in the May~6 data set
(Fig.~\ref{fig.fe12-fe13-may6}) is that the range of density variation of
\ion{Fe}{xii} is greater than that for \ion{Fe}{xiii}. 

Can the differences be due to a systematic error in the relative
calibration of the EIS instrument? By artificially varying the instrument
sensitivity at 202~\AA\ one finds that the densities from
\ion{Fe}{xii} \lam186.88/\lam195.12 and \ion{Fe}{xiii}
\lam196.54/\lam202.04 from the May~3 data set can be brought into line
if the sensitivity at 202~\AA\
sensitivity is increased by 35~\%\ (i.e., the measured DN value at
202~\AA\ converts to a smaller line intensity, thus increasing the
line ratio and the derived density). However such a large change would
also mean that the sensitivity at the nearby wavelength of 203.8~\AA\
must also be significantly increased, which would then spoil the
excellent agreement between the \ion{Fe}{xiii} \lam196.54/\lam202.04
and \lam203.82/\lam202.04 ratios seen in Fig.~\ref{fig.fe13-may6}. A
detailed study of insensitive line ratios around 190--205~\AA\ will be
required to confirm that the EIS relative calibration is sound, but we
believe a discrepancy of 35~\% is highly unlikely.

Ruling out a calibration problem, we can now ask: should \ion{Fe}{xii}
and \ion{Fe}{xiii} actually yield the same 
density values? Even if the two source regions for the emitting lines do
not have the same densities, an assumption of constant pressure  would
only lead to discrepancies of around 
0.1~dex (although the higher density this implies for the cooler
\ion{Fe}{xii} ion is consistent with the measurements). A key issue to
consider is whether the discrepancies could be caused through the
existence of multiple density components along the line of sight: an implicit assumption when  generating the plots is that the
plasma observed has a fixed density for the line of sight
corresponding to each pixel. This is generally not the case in
the corona. Even if a coronal loop, for example, has a fixed density
at the location being considered, the loop is sitting in a coronal
background that could have a varying density with height (in the case
of a hydrostatic equilibrium), or multiple densities due to a variety
of loop like structures along the line of sight. 
The measured density
then actually represents a weighted average of the different plasma
components \citep[e.g.,][]{doschek84}.

To look at this effect further, we consider the May~6 data set.
Consider the spatial region around Y-pixel 150 where the \ion{Fe}{xii}
and \ion{Fe}{xiii} densities become very similar. This region is seen in the
upper panels of Figs.~\ref{fig.fe12-may6} and \ref{fig.fe13-may6} to
be free of loops and thus correspond to the background corona. In the
nearby pixel regions 90--130 and 160--170 which are seen to
correspond to active region loops, the discrepancy between the
two ions sharply rises to 0.2--0.3~dex. Is this discrepancy due to the
background subtraction skewing the loop density measurements?

\begin{figure}[h]
\centerline{\epsfxsize=9cm\epsfbox{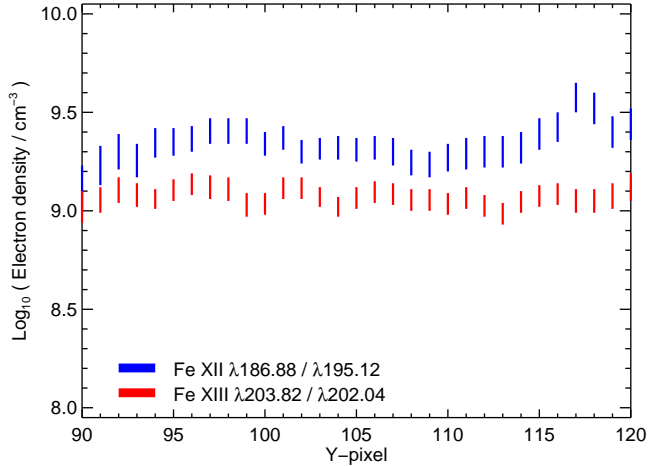}}
\caption{A comparison of densities derived from
  \ion{Fe}{xii} \lam186.88/\lam195.12 and \ion{Fe}{xiii}
  \lam203.82/\lam202.04 for the loop
  bundle located at Y-pixels 90--120 in the May~6 data set. A
  background level has been subtracted from each of the line
  intensities, thus leading to different densities than those shown in
  Figs.~\ref{fig.fe12-fe13-may6}.}
\label{fig.subtract-dens}
\end{figure}

Fig.~\ref{fig.subtract-dens} shows the densities derived from
\ion{Fe}{xii} \lam186.88/\lam195.12 and \ion{Fe}{xiii}
\lam203.82/\lam202.04 after a background contribution to the line
intensities has been subtracted. For each line, the background
intensity was calculated by averaging the intensities in Y-pixel
region 132 to 143. These background intensities were then subtracted
from the intensities in the `loop' region (Y-pixels 90--120), and the
densities re-derived. Comparing with Fig.~\ref{fig.fe12-fe13-may6} the
background subtractions have resulted in higher densities across the
loop region by about 0.2~dex, but the discrepancy between
\ion{Fe}{xii} and \ion{Fe}{xiii} has not been resolved. Considering
the central part of the loop region (Y-pixels 95--105) the average
difference between \ion{Fe}{xii} and \ion{Fe}{xiii} was 0.28~dex
before background subtraction, and becomes 0.27~dex after
subtraction. In summary, then, background subtraction does not affect
the \ion{Fe}{xii}--\ion{Fe}{xiii} discrepancy in this case.

The May~3 data set shows a much more complex plasma and so
considerations of background subtraction on the density values are
more difficult. However, the density in the Y-pixel region 60--80 is so
high that the emission from this region will overwhelm any contribution
from the background and thus we can speculate that it is a relatively
`pure' measurement of the region's density. Despite this the
discrepancy between the two ions' density values is around
0.4--0.5~dex. It is difficult to envisage any physical circumstances
that could result in such a discrepancy. We also note the discrepancy
over the Y-pixel region 100--175 is consistently at the 0.4--0.5~dex
level even though there is a wide range of plasma structures in this
region (see the upper panels of Figs.~\ref{fig.fe12-may3} and
\ref{fig.fe13-may3}). Therefore we
believe these discrepancies are largely due to 
atomic physics parameters for the two ions, rather than physical
properties of the plasma. 

A further check that can be performed on the diagnostics is to use the
derived densities to estimate the column depth of the emitting region
at each pixel. Comparisons of column depths from the
two ions with the observed sizes of solar structures potentially allow
a statement to be made about which  ion is yielding the most accurate
density values. This issue will be discussed in a following paper.

\section{Conclusions}

An active region was observed by \emph{Hinode}/EIS on 2007 May 3 and
May 6, and coronal density values have been derived from emission line
ratios of \ion{Fe}{xii} and \ion{Fe}{xiii} ratios that are the most
sensitive in the EIS wavelength bands. Methods for automatically
fitting the emission lines at each pixel in the spectral images  have
been described and line blending discussed. The derived densities are
presented in Figs.~\ref{fig.fe12-may6}--\ref{fig.fe12-fe13-may3} and
the following conclusions are drawn.

\begin{itemize}
\item In active region data sets the precision of the density
  measurements from the four \ion{Fe}{xii} and \ion{Fe}{xiii} can be
  as high as $\pm$5~\%\ in individual spatial pixels -- a considerable improvement over all previous
  solar spectrometers. 
\item The high precision of the measurements reveals significant
  discrepancies between the four ratios, thus implying the
  \emph{accuracy} of the measurements is only around a factor 2.
\item For \ion{Fe}{xii}, the \lam196.64/\lam195.12 ratio always yields
  higher densities than \lam186.88/\lam195.12 by up to 0.4~dex in $\log\,N_{\rm
  e}$. This is believed to be due to atomic data effects
  and/or instrumental effects, rather than a real physical effect.
\item For \ion{Fe}{xiii}, the \lam203.82/\lam202.04 ratio is in very good
  agreement with \lam196.64/\lam202.04 for densities less than $\log\,N_{\rm
  e}=10.0$. Above this density \lam203.82/\lam202.04 yields higher
  densities which is most likely due to an inaccuracy in the
  theoretical high-density limit for the ratio.
\item Comparing \ion{Fe}{xii} and \ion{Fe}{xiii}, the \ion{Fe}{xii}
  densities are higher than the \ion{Fe}{xiii} densities at almost all
  spatial locations, with the discrepancy being largest (up to
  0.5~dex) at high densities. This is most likely due to the atomic
  data for the ions, but no indication can be given here on which
  ion's densities are most accurate. 
\item The May~6 data set shows a greater
  range of density variation for \ion{Fe}{xii} than for \ion{Fe}{xiii}
  which may be due to real physical differences for the \ion{Fe}{xii} and \ion{Fe}{xiii}
  emitting regions.
\end{itemize}

In addition to the above a number of instrumental and data analysis
problems have been 
identified in this work

\begin{itemize}
\item The EIS spectra are tilted relative to the CCD's axis in the SW
  band with a slope of $-0.0792$~\AA/pixel, an effect due to a slight
  misalignment of the grating relative to the CCD. This can have a
  significant effect on derived densities and must be corrected for if
  the emission lines are separated by more than a few angstroms.
\item It is necessary to account for the \ion{Fe}{xii} line at
  195.18~\AA\ when interpreting the intensity of the strong
  \ion{Fe}{xii} \lam195.12 line at densities $\ge 10^{10}$~cm$^{-3}$.
\item \ion{Fe}{viii} \lam196.65 contributes to \ion{Fe}{xii}
  \lam196.64 in regions where \ion{Fe}{viii} emission is enhanced. The
  size of the contribution can be estimated by measuring
  \ion{Fe}{viii} \lam194.66, however the current CHIANTI model for
  \ion{Fe}{viii} appears to under-estimate the  contribution.
\item \ion{Fe}{xii}
  \lam186.88 is blended with \ion{S}{xi} \lam186.84, but the
  contribution of this line is found to be mostly at the level of $\le
  10$~\%, and has only a small effect on derived \ion{Fe}{xii}
  densities.
\item The rest wavelengths contained in CHIANTI for some of the
  \ion{Fe}{xii} and \ion{Fe}{xiii} lines are not correct. We find
  new wavelengths of $196.518\pm 0.003$~\AA\ for \ion{Fe}{xiii} \lam196.54, and
  $196.647\pm 0.003$~\AA\ for \ion{Fe}{xii} \lam196.64.
\end{itemize}

The discrepancies between the densities derived from the different
\ion{Fe}{xii}  and \ion{Fe}{xiii} diagnostics are a serious impediment to using
the densities for science analysis. A further study into the effects
of the discrepancies on plasma column depths and filling
factors will be presented in a following paper, which may shed light
on which ion yields the most accurate density values. We believe the
discrepancies are largely due to inaccurate atomic parameters for one
or both of the ions and so
new investigations into these parameters are urgently required in
order that scientists may make best use 
of the high quality EIS spectra.

\begin{acknowledgements}
We acknowledge F.P.~Keenan, G.~ Del Zanna and the anonymous referee
for useful comments on the manuscript. Hinode is a Japanese mission developed and launched by
ISAS/JAXA, with NAOJ as domestic partner and NASA and
STFC (UK) as international partners. It is operated by
these agencies in co-operation with ESA and NSC (Norway).
\end{acknowledgements}

\Online

\begin{appendix}

\section{An investigation of the Fe\,XII 195.12 blend}\label{app.195}

The strong \ion{Fe}{xii} \lam195.12 line has been known to have a
significant blending component due to the $3s^23p^3$ $^2D_{3/2}$ --
$3s^23p^2(^1D)3d$ $^2D_{3/2}$ \ion{Fe}{xii} transition 
since the work of \citet{binello01}, who placed the latter transition
at 195.13~\AA. The wavelength was revised to 195.179~\AA\ by
\citet{delzanna05} based on unpublished work by 
B.C.~Fawcett (G.~Del Zanna, private communication, 2008).
In this section we
investigate some of the properties of the fit parameters from our two
Gaussian fits. We do this by making comparisons with a 1 Gaussian fit
to the 195.12~\AA\ feature using the same wavelength pixels as marked
in Fig.~\ref{fig.195-example}.

\begin{figure}[h]
\centerline{\epsfxsize=7cm\epsfbox{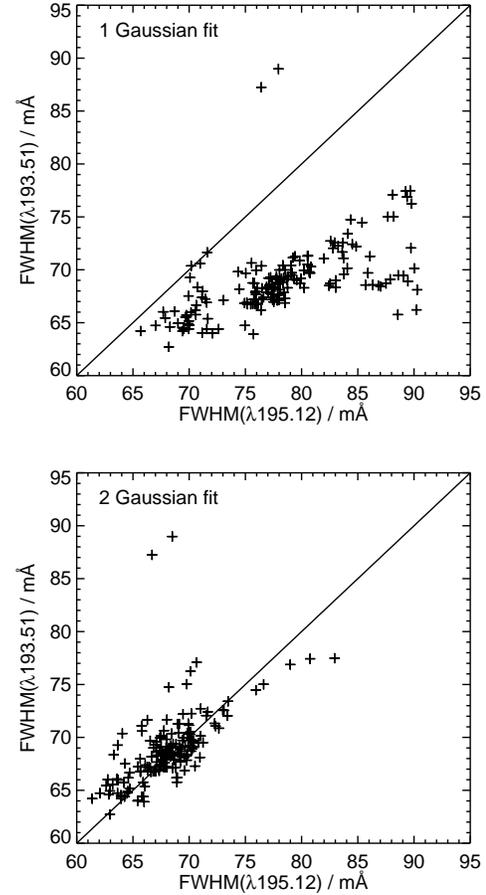}}
\caption{A comparison of the full widths at half maxima (FWHM) of
  Gaussian fits to the \ion{Fe}{xii} \lam195.12 and \lam193.51 lines
  in the May~3 data set. The upper panel shows the results when a
  single Gaussian fit is applied to \lam195.12, the lower panel shows
  the results when a two Gaussian fit is applied in order to account
  for the blending \lam195.18 line.}
\label{fig.fe12-width}
\end{figure}

Firstly we compare the line widths of \lam195.12. It has been
noted \citep{doschek07a} that this line is broader than nearby
\ion{Fe}{xii} \lam193.51, and \lam195.18 was suggested as a possible
cause of this \citep{young07b}. In Fig.~\ref{fig.fe12-width} we
compare the widths of 
\lam193.51 and \lam195.12 measured from the May~3 data set when
\lam195.12 is fit with a single 
Gaussian and when it is fit with a double Gaussian. The width of
\lam195.12 is clearly seen to be larger than that of \lam193.51 with
the one Gaussian fit, but they come into excellent agreement when a two
Gaussian fit is used. This demonstrates that \lam195.18 is reponsible
for the broader width of \lam195.12 noted by \citet{doschek07a}.

\begin{figure}[h]
\centerline{\epsfxsize=9cm\epsfbox{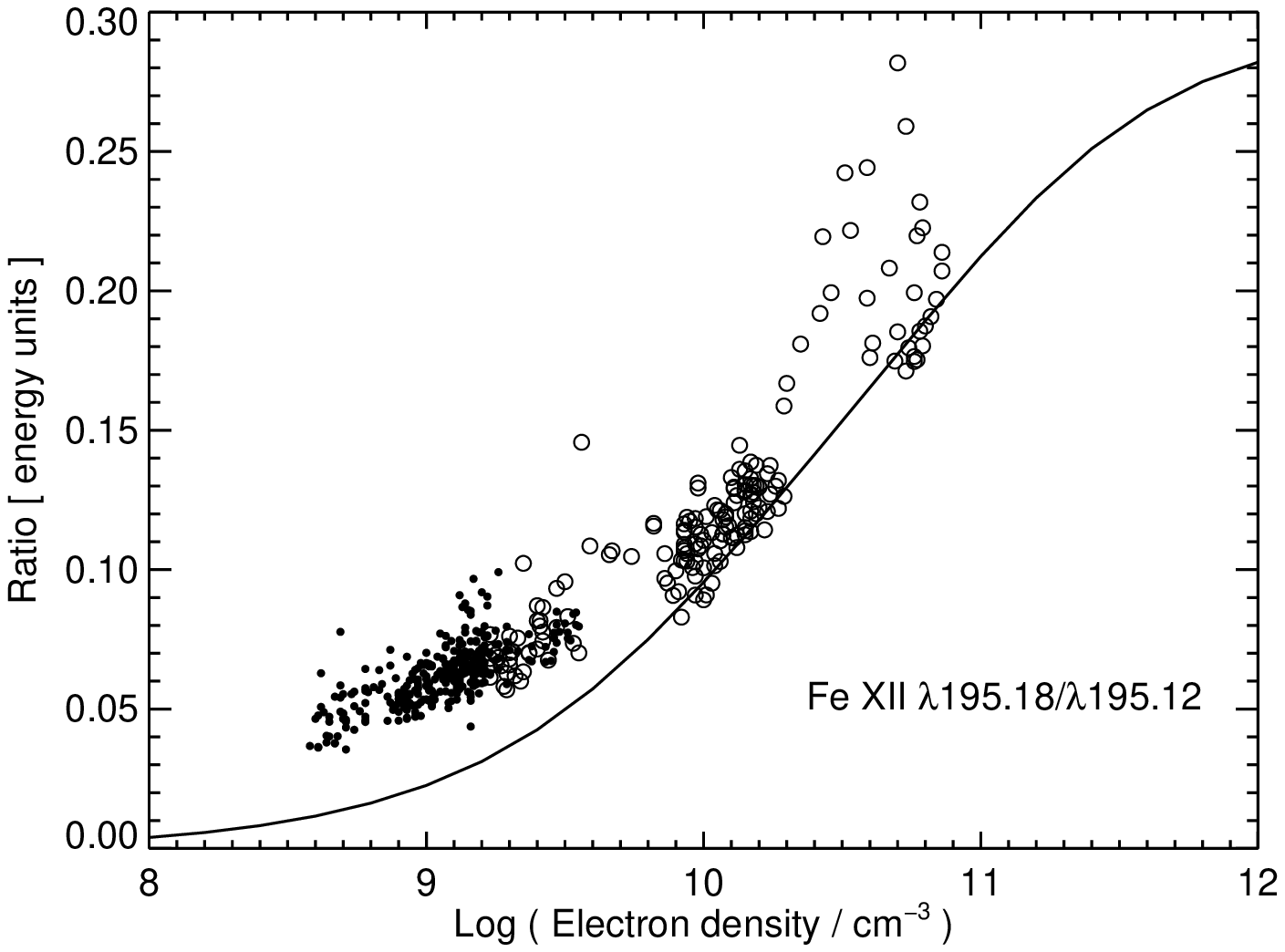}}
\caption{A comparison of the measured variation of the \ion{Fe}{xii}
  \lam195.18/\lam195.12 intensity ratio with density 
  compared to the predictions from the
  CHIANTI database. The measured densities are obtained from the
  \ion{Fe}{xii} \lam186.88/\lam195.12 ratio. Small filled circles denote
  measurements from the May~6 data set, while larger, open circles
  denote measurements 
  from the May~3 data set.}
\label{fig.195-ratio}
\end{figure}

A second issue related to \lam195.18 that can be investigated
is the density sensitivity of the
\lam195.18/\lam195.12 ratio. Fig.~\ref{fig.195-ratio} shows the
predicted variation of the ratio from CHIANTI where \lam195.18 is seen
to become relatively stronger at high densities. Overplotted on this
figure are observed ratios and densities from the May~6 and May~3
data sets. The densities are those obtained from the \ion{Fe}{xii}
\lam186.88/\lam195.12 ratio. For densities $\ge 10^{10}$~cm$^{-3}$
agreement is quite good, although there are a number of points
significantly above the theoretical curve which are likely due to the
\lam195.18 intensity being over-estimated by the two Gaussian fits.

At low densities, the observed ratios are all above the theoretical curve
although the shape of the curve is reproduced quite well. This may be
due to a systematic over-estimate of the \lam195.18 intensity in the
fits due again to an unaccounted-for line.

\section{The Fe\,XII triplet 192.39, 193.51, 195.12}\label{app.fe12}

The $3s^23p^3$ $^4S_{3/2}$ -- $3s^23p^2(^3P)3d$ $^4P_{5/2,3/2,1/2}$
transitions yield the three \ion{Fe}{xii} transitions at 195.12,
193.51 and 192.39~\AA, respectively. Their theoretical intensity
ratios are almost completely insensitive to temperature and density
and CHIANTI gives the values 1.0:0.67:0.32, respectively. All three
lines are very strong in 
the EIS spectra \citep{young07b} with \lam195.12 the strongest and so
this line is most commonly observed with EIS. However, in active
region observations \lam195.12 can be saturated making it difficult
to use. In such circumstances it may be necessary to use \lam192.39 or
\lam193.51 if they are observed with the EIS study as they are less
likely to be saturated. In this section the relative intensities as
measured from the May~3 and May~6 data sets are presented and compared
with the 
CHIANTI predictions.

The ratios of the three lines will clearly be affected by blending
with other species, and the contribution of \lam195.18 to the feature
at 195.12~\AA\ was 
highlighted in Appendix~\ref{app.195} which demonstrated that \lam195.18
is responsible for the enhanced broadening of the 195.12~\AA\ feature
when it is fit with a single Gaussian. Inspection of the line widths
of \lam192.39 and \lam193.51 reveals no anomalies consistent with a
lack of blending for these lines.

\begin{figure}[h]
\centerline{\epsfxsize=9cm\epsfbox{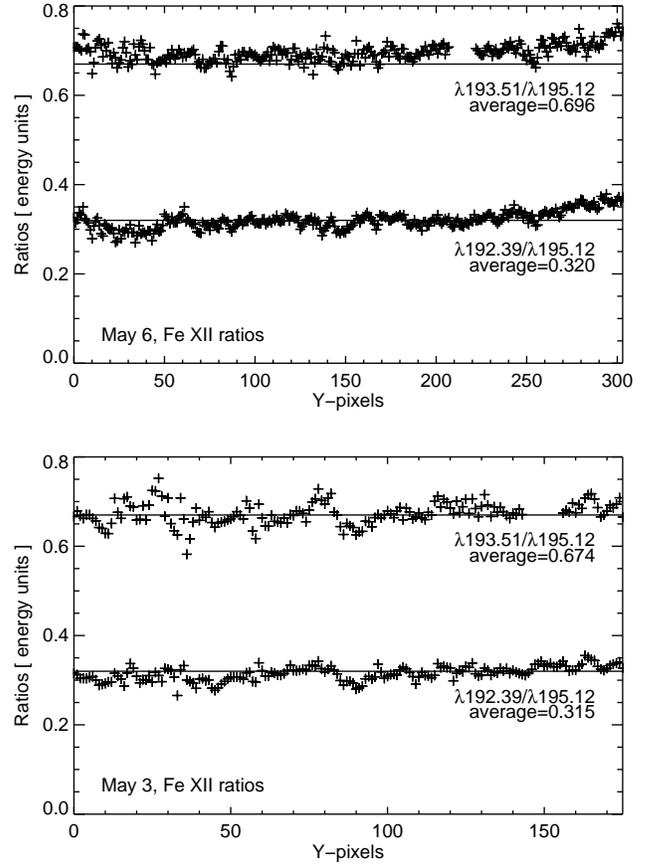}}
\caption{Comparison of measured \ion{Fe}{xii} \lam193.51/\lam195.12
  and \lam192.39/\lam195.12 intensity ratios against the 
  predictions from CHIANTI from the May~6 (upper panel) and May~3
  (lower panel) data sets. The solid horizontal lines in each panel
  show the line ratio values predicted from CHIANTI, and crosses
  denote the measured ratio values. The average of the measured values
  is shown for each of the ratios and data sets.}
\label{fig.fe12-ratios}
\end{figure}

Fig.~\ref{fig.fe12-ratios} shows the \lam192.39/\lam195.12 and
\lam193.51/\lam195.12 intensity ratios from the May~3 and May~6
data sets (lower and upper panels, respectively). Despite the large
variation in density and intensity for 
the data sets, the ratios show excellent agreement with the CHIANTI
theoretical values. The only discrepancy is that the
\lam193.51/\lam195.12 ratio in the May~6 data set is noticeably above
the CHIANTI prediction and the May~3 values, and no explanation can be
given for this effect.
Another effect that can be seen in both ratios is that the
measurements rise slightly as the Y-pixel increases. This can be
explained if the instrument sensitivity is slightly different towards
the top of the EIS slit than at the bottom, becoming more sensitive at
short wavelengths at the top. The laboratory measurements performed by
\citet{seely04} showed the shape of the multilayer coating
reflectivity varying with the position of the light source in the
aperture, and thus the \ion{Fe}{xii} ratios could be revealing this effect.

In summary, though, the \lam192.39/\lam195.12 and
\lam193.51/\lam195.12 ratios show excellent agreement with theoretical
predictions and thus either \lam192.39 or \lam193.51 can be used in
place of \lam195.12 when forming density diagnostics.

\section{Correcting for spatial offsets between emission lines}\label{sect.offsets}

The spectral image formed on the SW CCD is marginally tilted relative
to the CCD axes such that the same spatial feature is observed to be
slightly lower on the CCD in long wavelength lines compared to short
wavelength lines. The effect is due to a misaligment of the grating
relative to the CCD and we refer to it here as the ``grating tilt''. The best way to measure the tilt is to identify emission
lines formed at the same temperature and which are density insensitive
relative to each other. If a compact spatial feature is observed in
such lines, then the offset can be estimated by co-aligning the images
in the two lines. For the EIS SW band, however, there are few emission
line pairs that are insensitive and span a wide enough wavelength
range to be useful. In addition, most of the lines are formed in the
corona where 
spatial features are more diffuse than in the transition region,
making alignment more difficult. A further complication is that the tilt is
small (a one spatial pixel offset over 12.6~\AA\ is found here),
meaning for most line pair comparisons the offset is on the sub-pixel
level.

For the present work the focus is on the density sensitive lines of
\ion{Fe}{xii} and \ion{Fe}{xiii} which are found between 186.8 and
203.8~\AA. We thus seek to identify the grating tilt over this
region. We make use of the May~3 and May~6 data sets that have been
used for the density analysis, and four line pairs are
considered. \ion{Fe}{viii} images
often show compact features that are very suitable for co-alignment
\citep{young07a}, and the 
\lam185.21 and 
\lam194.66 lines are used here. They have a wide wavelength separation
and only a small 
density sensitivity, however \lam185.21 is blended with a
\ion{Ni}{xvi} line and so care has to be taken to avoid comparing spatial
features where there is clearly hot emission. 

The remaining three ratios are all from \ion{Fe}{xii}.  \ion{Fe}{xii}
\lam186.88/\lam196.64 involves two of the lines from  the density
analysis. CHIANTI predicts the intensity 
ratio to vary from 0.21 to 0.27 over the density range
$10^{8}$--$10^{12}$~cm$^{-3}$ and so the ratio is relatively
insensitive to density. \ion{Fe}{xii}
\lam192.39/\lam195.12 involves two very strong lines that are
insensitive to density (Appendix~\ref{app.fe12}), however the wavelength separation is
small. The last ratio is \lam203.73/\lam195.12 which, although it
shows significant density sensitivity, was deemed the best means of
estimating the grating tilt out to the spectral feature at
203.8~\AA. \lam203.73 is measured through a multi-Gaussian fit to
\ion{Fe}{xiii} \lam203.82 (Sect.~\ref{sect.203}).

The method for finding the spatial offset between two lines is as
follows. Each of the lines was fit with a Gaussian as described in
Sect.~\ref{sect.lines}, yielding intensity images in each of the
lines. A compact feature was then identified in particular image
columns, and the intensity profiles along these columns
extracted.  One of the two  columns
was treated as a reference and, for the other, the intensity values
were adjusted to simulate moving the image up and down on the
detector. The intensity for this comparison column was re-calculated as
\begin{equation}\label{eq.offset}
I_{\rm new}(i) = \left\{ \begin{array}{l@{\quad}l}
                (1-\alpha) I(i) + \alpha I(i+1) & \alpha > 0\\
                \\
                (1-|\alpha|) I(i) + |\alpha| I(i-1) & \alpha < 0 \\
                \end{array}
                \right. 
\end{equation}
where $\alpha$ is varied from $-1$ to $+1$ in 0.1 intervals. There are
thus 21 columns to be compared with the reference column. From the
reference column, the portion containing the compact
spatial feature was extracted and compared with the 21 comparison columns
to determine which gave the best spatial match. This was done by first
normalising the extracted columns to the maximum value in the
extracted region, and then calculating the quantity:
\begin{equation}
w^2 = \sum_i (I^*_{\rm ref}(i) - I^*_{\rm new}(i))^2
\end{equation}
where $I^*_{\rm ref}$ is the normalised intensity of the reference
wavelength, and $I^*_{\rm new}$ is the normalised comparison intensity (Eq.~\ref{eq.offset}).
The best match to the reference column was determined to be the
comparison column with the lowest $w^2$ value.

The May~6 data set had limited value for estimating the spatial
offsets as \ion{Fe}{xii} does not demonstrate any compact spatial
features (e.g., the upper panel of Fig.~\ref{fig.fe12-may3}). However,
there is a bright, compact feature seen in \ion{Fe}{viii} (lower panel
of Fig.~\ref{fig.fe12-may3}). The offset from the \lam185.21, \lam194.66
comparison was found to be $0.7$~pixels.

The May~3 data set showed a number of compact features at different
temperatures however the \ion{Fe}{viii} comparison was compromised by the
blending of \lam185.21 and so the offset from the May~6 dataset was
preferred. Comparing several features, the offset between \ion{Fe}{xii}
\lam186.88 and \lam196.64 was found to vary from 0.7 to 0.8 pixels --
we assume 0.75 here. For \lam195.12 and \lam203.72 the offset was 0.8
pixels, and for \lam192.39 and \lam195.12 it was 0.2 pixels.

Assuming that the grating tilt is linear, the offset for a given
wavelength is written as
\begin{equation}
y=m(\lambda - 195.12)
\end{equation}
where the offset is assumed to be zero at 195.12~\AA. Computing the
tilt from the four ratios and taking the average, we find
$m=-0.0792$~pixels/\AA.

\begin{figure}[h]
\centerline{\epsfxsize=7cm\epsfbox{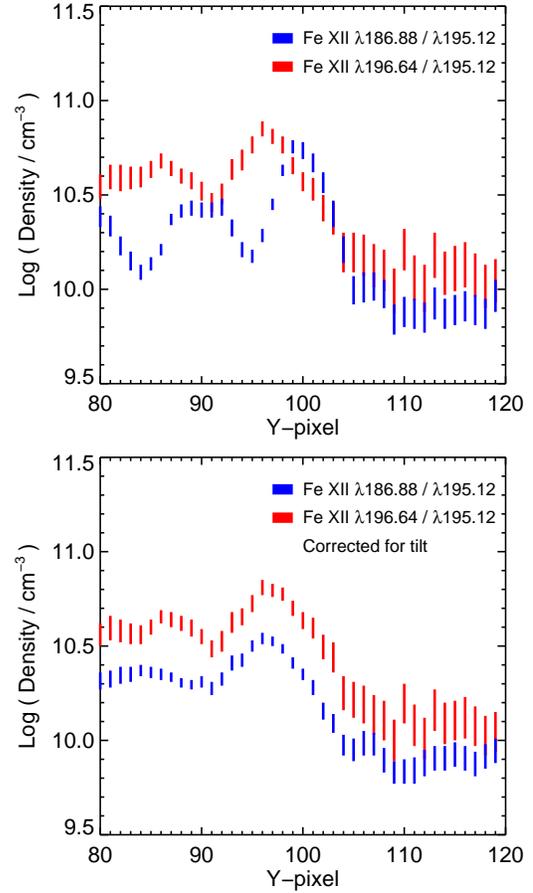}}
\caption{A comparison of densities obtained from the \ion{Fe}{xii}
  \lam186.88/\lam195.12 and \lam196.64/\lam195.12 ratios measured from
  the May~3 data set. A section of a data column from X-pixel 25 (see upper panel
  of Fig.~\ref{fig.fe12-may3} for reference) is shown where is found a
  compact high density feature. The upper panel shows the densities
  obtained when the grating tilt is not corrected for, and the lower
  panel the results when the tilt is corrected for. Note that the
  discrepancy between the density peaks in the upper panel is resolved
  in the lower panel.}
\label{fig.fe12-offset}
\end{figure}

To illustrate the consequence of the grating tilt on the derived
densities, we consider a small brightening that appears at X-pixel 25
and Y-pixel 98 in the May~3 data set (see
Fig.~\ref{fig.fe12-may3}). Fig.~\ref{fig.fe12-offset} compares the
densities derived from the two \ion{Fe}{xii} density diagnostics in
the vicinity of this brightening. The top panel shows the densities
when no offsets are applied, the bottom panel shows the densities when
the offsets have been applied. It is clearly seen that the peaks in
density around pixels 96--100 do not coincide in the top panel, but
do coincide in the bottom panel. The discrepancy in the top panel is
striking and greater than would be expected simply from the sub-pixel
spatial offsets derived for the lines. This is because taking the
ratios of lines amplifies the effect of the intensity offsets. 

Finally we note that there is some uncertainty in deriving the
grating tilt due to the difficulty of co-aligning data at the
sub-pixel level. This thus represents a possible systematic error in
the derived density measurements that will be greatest for those lines
with a large wavelength separation. It may affect density comparisons
between ratios for small spatial features, but not the broad trends
highlighted in Sect.~\ref{sect.discussion}.

\end{appendix}

\end{document}